\documentclass[fleqn,usenatbib]{mnras}

\DeclareRobustCommand{\VAN}[3]{#2}
\let\VANthebibliography\thebibliography
\def\thebibliography{\DeclareRobustCommand{\VAN}[3]{##3}\VANthebibliography}

\usepackage{graphicx}	
\usepackage{amsmath}	
\usepackage{amssymb}	
\usepackage{subcaption}
\title[NIR metallicity map of the LMC]{The VMC survey -- XLIV: Mapping metallicity trends in the Large Magellanic Cloud using near-infrared passbands}

\author[S. Choudhury et al.]{Samyaday Choudhury$^{1,2}$\thanks{E-mail: samyaday.choudhury@gmail.com},
Richard de Grijs$^{1,2}$, Kenji Bekki$^{3}$, Maria-Rosa L. Cioni$^{4}$, Valentin D. Ivanov$^{5}$,
\newauthor {Jacco Th. van Loon$^{6}$}, Amy E. Miller$^{4,1}$, Florian Niederhofer$^{4}$, Joana M. Oliveira$^{6}$, Vincenzo Ripepi$^{7}$, 
\newauthor {Ning-Chen Sun$^{8}$ and Smitha Subramanian$^{9}$}\\
$^{1}$ Department of Physics and Astronomy, Macquarie University, Balaclava Road, Sydney, NSW 2109, Australia\\
$^{2}$ Research Centre for Astronomy, Astrophysics and Astrophotonics, Macquarie University, Balaclava Road, Sydney, NSW 2109, Australia\\
$^{3}$ ICRAR, M468, University of Western Australia, 35 Stirling Hwy, 6009 Crawley, Western Australia, Australia\\
$^{4}$ Leibniz-Institut f\"{u}r Astrophysik Potsdam (AIP), An der Sternwarte 16, D-14482 Potsdam, Germany\\
$^{5}$ European Southern Observatory, Karl-Schwarzschild-Str. 2, D-85748 Garching bei M\"{u}nchen, Germany\\
$^{6}$ Lennard-Jones Laboratories, Keele University, ST5 5BG, UK\\
$^{7}$ INAF--Osservatorio Astronomico di Capodimonte, via Moiariello 16, I-80131, Naples, Italy\\
$^{8}$ Department of Physics and Astronomy, University of Sheffield, Hicks Building, Hounsfield Road, Sheffield S3 7RH, UK \\
$^{9}$ Indian Institute of Astrophysics, II Block, Koramangala, Bengaluru 560 034, India\\
}

\date{Accepted XXX. Received YYY; in original form ZZZ}

\pubyear{2015}

\begin{document}
\label{firstpage}
\pagerange{\pageref{firstpage}--\pageref{lastpage}}
\maketitle

\begin{abstract}
We have derived high-spatial-resolution metallicity maps covering
$\sim$105~deg$^2$ across the Large Magellanic Cloud (LMC) using
near-infrared passbands from the VISTA Survey of the Magellanic
Clouds. We attempt to understand the metallicity distribution and
gradients of the LMC up to a radius of $\sim$ 6~kpc. We identify red
giant branch (RGB) stars in spatially distinct $Y, (Y-K_{\rm s})$
colour--magnitude diagrams. In any of our selected subregions, the RGB
slope is used as an indicator of the average metallicity, based on
calibration to metallicity using spectroscopic data. The mean LMC
metallicity is [Fe/H] = $-$0.42~dex ($\sigma$[Fe/H] = 0.04~dex). We
find the bar to be mildly metal-rich compared with the outer disc,
showing evidence of a shallow gradient in metallicity ($-0.008 \pm
0.001$ dex kpc$^{-1}$) from the galaxy's centre to a radius of
6~kpc. Our results suggest that the LMC's stellar bar is chemically
similar to the bars found in large spiral galaxies. The LMC's radial
metallicity gradient is asymmetric. It is metal-poor and flatter
towards the southwest, in the direction of the Bridge. This hints at
mixing and/or distortion of the spatial metallicity distribution,
presumably caused by tidal interactions between the Magellanic Clouds.
\end{abstract}

\begin{keywords}
stars: abundances -- Hertzsprung-Russell and colour-magnitude diagrams -- galaxies: abundanes -- Local Group -- Magellanic Clouds 
\end{keywords}


\section{Introduction}

The Magellanic Clouds (MCs), comprising the Large Magellanic Cloud
(LMC) and the Small Magellanic Cloud (SMC), are among the nearest
dwarf satellite galaxies \citep[at distances of, respectively,
  $\sim$50 and 61~kpc;][]{deGrijs+2014AJ,deGrijs+2015AJ}. Together
with the Milky Way (MW), they represent the closest example of an
interacting system of galaxies
(\citealt{Murai&Fujimoto1980theMagStream, Tanaka1981theMagStream,
  Fujimoto&Murai1984theMagStream, Gardiner+1994numsimul,
  Westerlund1997theMCs}). Studies using the {\sl Hubble Space
  Telescope} \citep{Kallivayalil+2006a,Kallivayalil+2006b,Kallivayalil+2013}
suggest that the MCs are experiencing their first infall towards the
MW. Simulations show that the origin of various dynamical features
seen in the outskirts and inter-cloud regions, e.g., the Magellanic
Bridge or Stream, may be caused by periodic interactions between the
Clouds \citep{Diaz&Bekki2011MNRASconstrain,Besla+2012MNRAStherole,Wang+2019MNRAS}.

 The SMC exhibits dynamically distorted features in its stellar
  populations because of interactions with the LMC
  \citep[e.g.][]{Nidever+2013, Smitha+2017MNRAS,
    Dalal+2019MNRAS,Florian+2021, Luri+2021A&A}. Several distinct
  (sub)structures have been discovered in the LMC's outskirts
  \citep[e.g.][]{Mackey+2016MNRAS, Mackey+2018ApJ, Dalal+2021MNRAS},
  some of which could have resulted from interactions with the
  SMC. Spatial mapping of the galaxies' metallicity distribution,
  including of any metallicity gradients (MGs), is an important tool
  to observe and understand the effects of the Clouds' evolution and
  interactions on their stellar chemical content.  In
  \citet[][henceforth C20]{C20}, we created SMC metallicity maps out
  to a radius of 4$^\circ$ from the galaxy's optical centre. We found
  evidence of asymmetric radial MGs in the SMC, which appeared flatter
  towards the LMC. This is a potential indicator of stellar mixing,
  likely caused by LMC--SMC interactions. Here, we check if similar
  effects are seen in the LMC.

The most accurate way to estimate metallicities is by means of
spectroscopic studies. Several studies have used Ca~{\sc ii}
  triplet (CaT) spectroscopy of red giant branch (RGB) stars to study
  star clusters and field regions in the LMC. Most studies were limited in
  both sample size (a few tens of clusters and a few hundred RGB
  stars) and spatial coverage. Either little or no evidence of radial
  MGs has been reported \citep[e.g.][]{Olszewski+1991AJspectro,
    Cole+2005AJspectroOfRGs, Grocholski+2006AJCaIItriplet,
    Carrera+2008AJ-CEH-LMC, Pompeia+2008A&Achemi,
    Lapenna+2012ApJtagging}. With recent large-area spectroscopic
surveys of the MCs, e.g., the Apache Point Observatory Galactic
Evolution Experiment's \citep[APOGEE;][]{Majewski+2017AJ} southern sky
survey \citep{Nidever+2020ApJ}, opportunities have opened up for
studying MGs, as well as of the individual elemental abundances and
chemo-kinematics in the MCs. \cite{Nidever+2020ApJ} reported
metallicities of 3600 RGB stars based on high-resolution $H$-band
spectra from the APOGEE survey. \cite{Skowron+2021ApJS} used the
APOGEE data of red giant (RG) stars from \cite{Nidever+2020ApJ} to
estimate an LMC MG of $-$0.026$\pm$0.002 dex deg$^{-1}$ out to around
8$^\circ$.

Since the LMC occupies a wide area across the sky, large photometric
surveys are helpful to estimate photometric metallicities and their
variation across the galaxy. Spectroscopic surveys like APOGEE
do not cover the MCs homogeneously to the same extent as large-area
photometric studies. Photometric estimates, although not as accurate
as spectroscopic ones, can help map relative metallicity variations
across a galaxy and estimate global properties including radial
MGs. \cite{Cioni2009A&Athemetallicity} estimated an LMC MG of
  $-0.047\pm0.003$ dex kpc$^{-1}$ out to $\sim$8~kpc using asymptotic
  giant branch (AGB) field stars. They used the C/M (Carbon to Oxygen)
  ratios of their AGB stars as a metallicity indicator. Although these
  authors covered a large area across the LMC, their indicators (AGB
  stars) and calibrators (RGB stars) were different, and the C/M ratio
  is potentially susceptible to age effects. Later,
  \cite{Feast+2010MNRAS} re-analysed the data of
  \cite{Cioni2009A&Athemetallicity} and confirmed a small gradient out
  to about 4~kpc.

\citet[][henceforth C16]{C16} estimated first-of-their-kind
metallicity maps for the LMC by combining large-area optical ($V$- and
$I$-band) photometric surveys and CaT spectroscopic data of RGB
stars. They estimated MGs out to a radius of 4~kpc: $-$0.049$\pm$0.002
dex kpc$^{-1}$ and $-$0.066$\pm$0.006 dex kpc$^{-1}$ based on
Magellanic Cloud Photometric Survey
\citep[MCPS;][]{Zaritsky+2004AJmcps} and Optical Gravitational Lensing
Experiment phase~III \citep[OGLE~III;][]{Udalski+2008AcAogleIII} data,
respectively. Recently, \cite{Grady+2021ApJ} used photometric data of
RGB stars from {\it Gaia} data release 2 (DR2) and combined them with
other photometric surveys, including the Wide-field Infrared Survey
Explorer \citep[WISE;][]{Wright+2010AJ}, the Two Micron All Sky Survey
\citep[2MASS;][]{Skrutskie+2006AJ} and APOGEE spectroscopic data, to
estimate metallicity maps of the MCs by employing machine-learning
techniques. They estimated a negative LMC MG of $-$0.048$\pm$0.001 dex
kpc$^{-1}$ out to 12~kpc, in agreement with
\cite{Cioni2009A&Athemetallicity} and C16.

 Here, we employ C16's technique to produce LMC metallicity maps
  using the near-infrared (NIR) VISTA Survey of the Magellanic Clouds
  \citep[VMC;][]{Cioni+2011A&A}. The VMC survey covers an approximately two to three times larger area
than the OGLE~III and MCPS surveys. We will thus be able to derive
metallicity maps and gradients out to $\sim$6.5~kpc, covering a much
larger area than C16.

The layout of this paper is as follows. In Section~2 we describe the
LMC VMC data. Section~3 contains our analysis of the RGB slope
estimation and its calibration to metallicity. We present our main
results, the NIR metallicity maps of the LMC, in Section~4. A
discussion is presented in Section~5. We summarise our conclusions in
Section~6.

\section{Data} 
The VMC survey is a uniform and homogeneous survey of the Magellanic
System in NIR passpands ($Y, J$ and $K_{\rm s}$) using the 4m VISTA
telescope \citep{Sutherland+2015} at La Silla Paranal Observatory,
Chile. It is one of the European Southern Observatory's (ESO) public
surveys. We direct readers to \cite{Cioni+2011A&A} for a detailed
description of the survey and its science goals. The VMC observations
began in 2009 and were completed in 2018. The survey covers about
$\sim$170 deg$^2$ of the Magellanic System using the VISTA infrared
camera \citep{Dalton+2006SPIE}. The area coverage is $\sim$105~deg$^2$, of which 42~deg$^2$, 20~deg$^2$ and 3~deg$^2$ cover the LMC,
SMC, and the Magellanic Bridge and Magellanic Stream,
respectively. The survey is deep, with total exposure times of
  2400~s, 2400~s and 9000~s in the $Y$, $J$ and $K_{\rm s}$ passbands,
  respectively. A single tile of VMC observation represents a mosaic
of six paw-print images in a given passband ($YJK_{\rm s}$). The
number of such tiles covering the LMC is 68. Each of these tiles
covers almost uniformly an area of 1.5 deg$^2$ \citep[by a minimum of
  two pixels;][]{Sutherland+2015} and their centres extend out to 6.5~kpc from the LMC's centre.

In this study, we use the point spread function (PSF) photometry
catalogue of the LMC. PSF photometry was performed following
  \cite{Rubele+2015}. The catalogue was calibrated using the VISTA
  photometric zero-points (v1.3). Similar to C20 (see their Section~2), we excluded sources with bad pixels and possible extended sources
by selecting stars with sharpness values between $-1$ and $+1$
\citep[see][their appendix]{Rubele+2015}. Overall, the $Y$- and
$K_{\rm s}$-band photometry has a typical uncertainty of $\leq$
0.15~mag, for stars brighter than $Y=$ 19~mag and $K_{\rm s}=$
18.5~mag. For our analysis we consider only the upper part of the RGB
in the ($Y$, $Y-K_{\rm s}$) colour--magnitude diagrams (CMDs),
provided that the stars are brighter than the above magnitude
limits. We only consider stars with photometric uncertainties $\le
0.15$~mag.


\section{Analysis}

\subsection{Estimation of the RGB slopes}
\begin{figure} 
\begin{center} 
\includegraphics[width=\columnwidth]{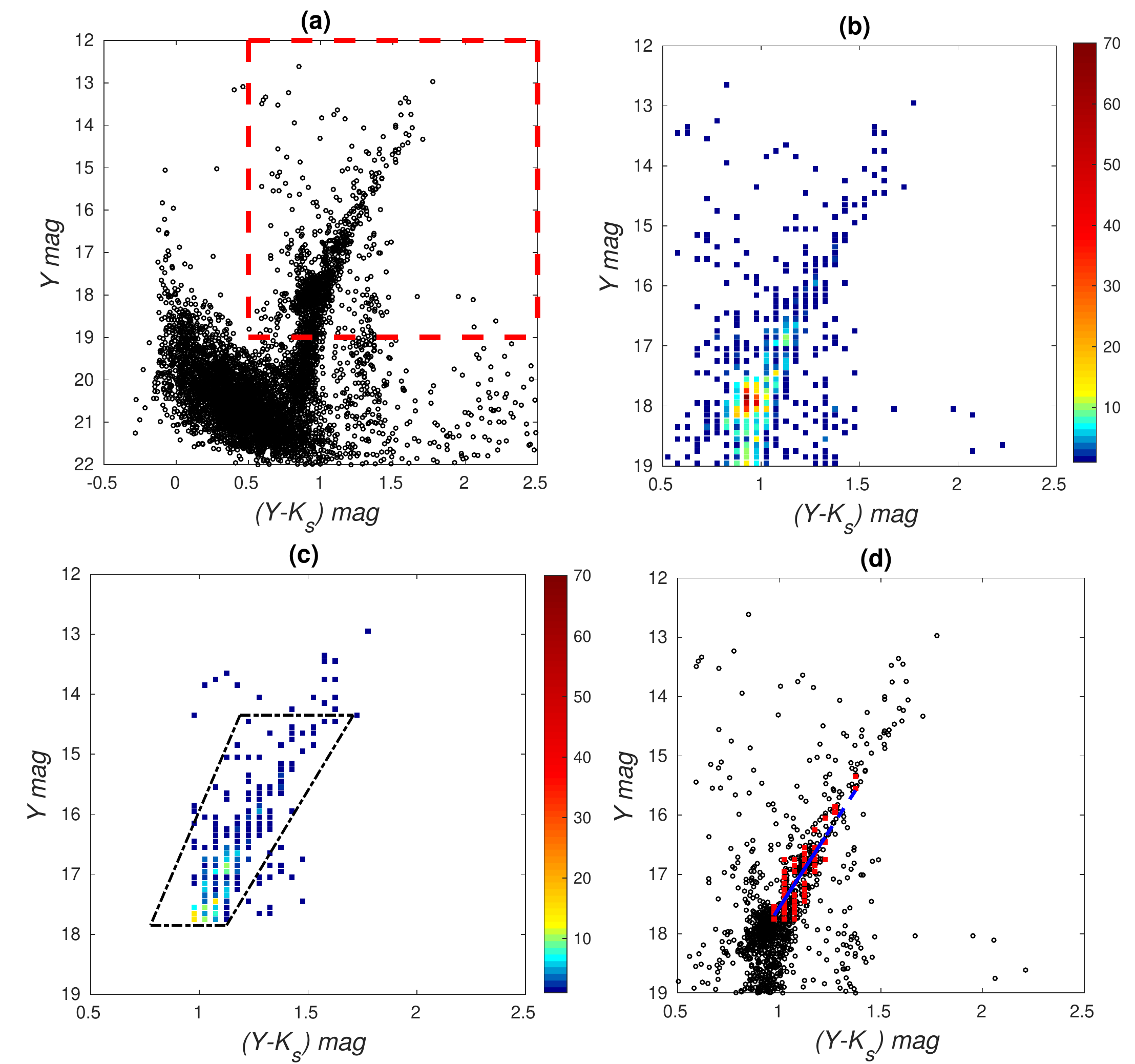}
\caption {(a) $Y$ versus $(Y-K_{\rm s})$ CMD of a 7.85$\times$8.10
  arcmin$^2$ LMC subregion at RA = $71.49^\circ$, Dec = $-70.28^\circ$
  in tile LMC 5$\_$2, containing $N=6351$ stars (black
    points). Stars in the rectangle (red dashed line) belong to the
  evolved part of the CMD. (b) Density diagram of the evolved part of
  the CMD, where the bins are colour-coded based on stellar numbers
  (see the colour bar). (c) Density diagram following the application of a
  colour--magnitude cut at the peak of the RC distribution. The
  trapezoid (black dashed lines) was defined to select the RGB region
  and remove the dominant MW contamination. (d) CMD of the subregion
  (black points) overplotted with bins containing $\ge$3 stars
  each (red filled squares). A linear fit to these bins representing
  the RGB is shown as the blue dashed line. The colour bars in (b) and
  (c) represent the numbers of stars in each colour--magnitude
  bin. The estimated parameters are: $|$slope$| = 5.27 \pm 0.48$ mag
  mag$^{-1}$, $r =0.86$ and $N_p = 44$.}
\label{fig:lmcrgbfit}
\end{center} 
\end{figure}

\begin{figure} 
\begin{center} 
\includegraphics[width=\columnwidth]{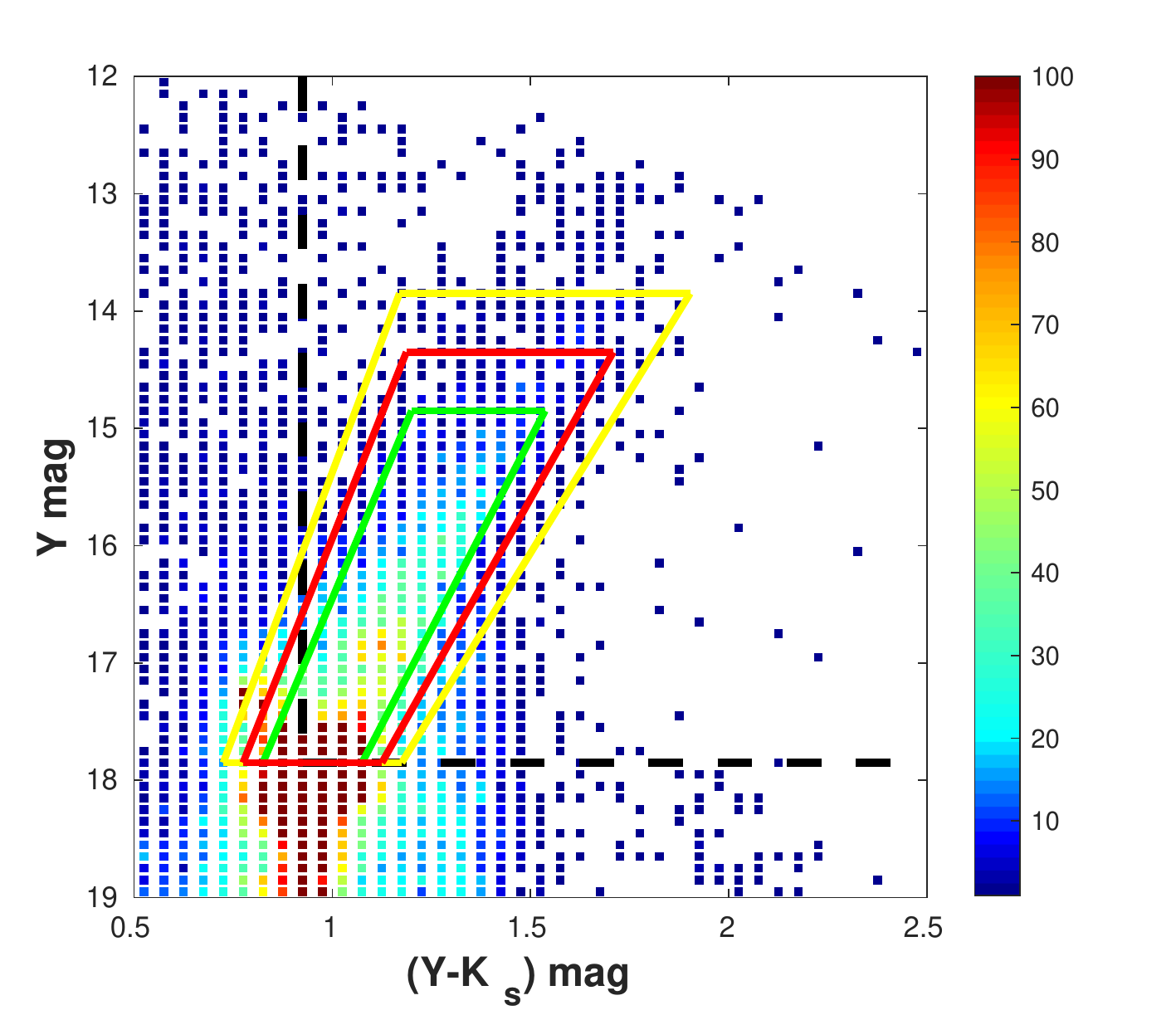}
\caption{Density CMD for six combined subregions selected from the
  central LMC and from the outer parts in four different directions
  (see text for details). The black dashed lines show the
  colour--magnitude cut at the peak of the RC distribution. Three
  different sizes of trapezoids are shown (yellow, red and green solid
  lines) to select the dominant RGB population and subtract MW
  contamination. All are defined with respect to the RC peak. We use
  the red trapezoid for our analysis. The colour bars have the same
  meaning as those in Fig.~\ref{fig:lmcrgbfit}.}
\label{fig:combocmd}
\end{center} 
\end{figure}
The LMC tiles are divided into 30 or more finer areas (referred
to as subregions throughout the text) based on stellar density. We
will briefly describe our method below. However, readers are directed
 to section~3 of C16 for a detailed description of the steps
adopted to estimate the RGB slope in the CMDs of the subregions. Their technique identifies the dominant  population and estimates its slope for field populations. A
salient feature of the technique is that we adopt the densest part of
the red clump (RC) in the relevant CMD as the base of the RGB. Since
the CMD loci of the RC and RGB stars are similarly affected by
reddening, the technique can be automated for the entire LMC,
irrespective of the prevailing reddening. Following C20, we selected
the two extreme VMC passbands, $Y$ and $K_{\rm s}$, for our
study. This affords us access to the longest colour baseline, and
ensures that we have access to the maximum observable effect of
metallicity on the upper RGB (the RGB segment brighter than the RC),
and hence on the RGB slope (for further explanation, see Section~3 of
C20). The main differences between this study and C20 are related
  to the removal of MW contamination.

 We define the estimated parameters for the subregions
  consistently with respect to our previous work. The total number of
stars in a subregion is denoted by $N$. We define $N_p$ as the number
of CMD bins (with the number of stars in each bin $\ge$3) representing
the RGB. These bins are fitted with a straight line and the slope
($|$slope$|$$\pm$$\sigma_{\rm slope}$) is estimated using
least-squares minimisation (using 3$\sigma$ clipping in a single
iteration). The correlation coefficient of the fit (also for other
fits in the remainder of the paper) is assessed by its absolute value,
$r$. We emphasise that, similarly to our previous studies, we are
interested in the relative variation in RGB slope (hence in
metallicity) among subregions rather than in absolute estimations for
individual subregions.

\begin{figure} 
\includegraphics[width=\columnwidth]{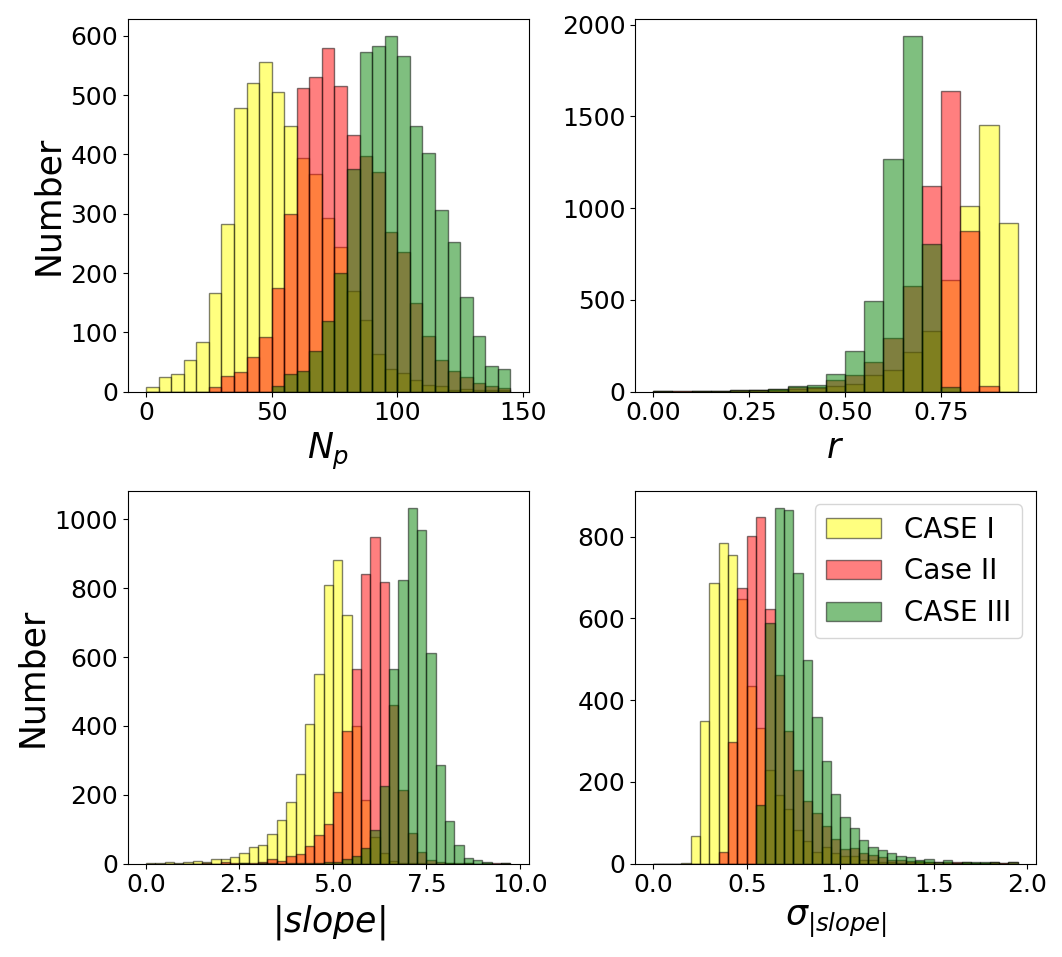}
\caption{Histograms of parameters ($N_p$, $|$slope$|$, $\sigma_{\rm
    slope}$ and $r$) estimated for the RGB adopting different sizes of
  the trapezoid considered to select the RGB region. Case II is
    for the final size adopted. Cases I and III are trapezoids
    covering smaller and larger areas, respectively, as shown in
    Fig.~\ref{fig:combocmd}. The histograms are shifted with respect
    to each other by small constant values along the $x$ axis to avoid
    overlap. Offsets are 25, $-$0.10, 1.0 and 0.15 for $N_p$, $r$
    $|$slope$|$ and $\sigma_{\rm slope}$, respectively.}
\label{fig:quad_hist}
\end{figure}

\begin{enumerate}
\item We begin by excluding the main sequence and isolating the
  evolved portion of the CMD (Fig.~\ref{fig:lmcrgbfit}a). Then, we
  construct a density distribution of the evolved portion to identify
  the locus of the RC. We adopt the peak in the RC's $(Y-K_{\rm s})$
  colour and $Y$ magnitude as the base of the RGB
  (Fig.~\ref{fig:lmcrgbfit}b) and apply a colour--magnitude cut at
  this peak (Fig.~\ref{fig:lmcrgbfit}c).
    
\item Next, we adopt a trapezoid to select the RGB region and
  discard MW contamination. The MW contamination is defined by two
  features \citep[see][]{Rubele+2012A&A}. The one bluer than the RGB
  and resembling a vertical bright strip in the vicinity of the RC
  distribution is removed by the colour and magnitude cut applied at
  the RC peak. The other feature is similar to a vertical strip in the
  redder part of the CMD, close to the RGB base. This is because the
  LMC is closer and and more metal-rich than the SMC. This choice of
  trapezoid proved to be better than the stepped block adopted in C20
  to remove MW contamination. We define the trapezoid with its base
  anchored to the peak colour and magnitude of the RC
  distribution. Hence, the RGB selection criteria are consistent
  across all subregions.
    
  Fig.~\ref{fig:combocmd} shows how we select the shape and size of
  the trapezoid. The shape is similar to that of region `K' adopted by
  \cite{Dalal+2019MNRAS} to select the RGB population from VMC $(K_{\rm
    s}, J-K_{\rm s})$ CMDs to estimate morphological maps. We start
  by creating a density diagram by combining CMDs of subregions from
  different locations within the LMC: central region -- tiles LMC
    7$\_$4 and 7$\_$5; outer regions in all four directions -- tiles
    LMC 2$\_$5, 5$\_$2, 6$\_$9 and 9$\_$3. We plot trapezoids of
  various sizes to sample the dominant part of the RGB population. We
  define the trapezoid carefully so as to avoid the redder MW
  contamination and avoid removing portions of the RGB. We check for
  systematic effects for three choices of size on our estimated
  parameters: see Fig.~\ref{fig:quad_hist}. We hardly find any
  difference in the distributions of the estimated parameters for any
  of the three different sizes adopted. For our analysis, we
  eventually proceed with the red trapezoid shown in
  Fig.~\ref{fig:combocmd}. We also verified that removing MW
    contamination from within the upper RGB region has a negligible
    effect on the resulting slope values. According to
    \cite{Dalal+2019MNRAS} only 6.4 per cent of stars in the upper RGB
    region are from the MW. This implies that the RGB is still the
    dominant population there. Decontamination using {\it Gaia} proper
    motions may lead to slope variations of $\pm$0.05--0.15, which is
    well within the 1$\sigma$ uncertainties.
\end{enumerate}

\begin{figure}
\centering
\includegraphics[width=\columnwidth]{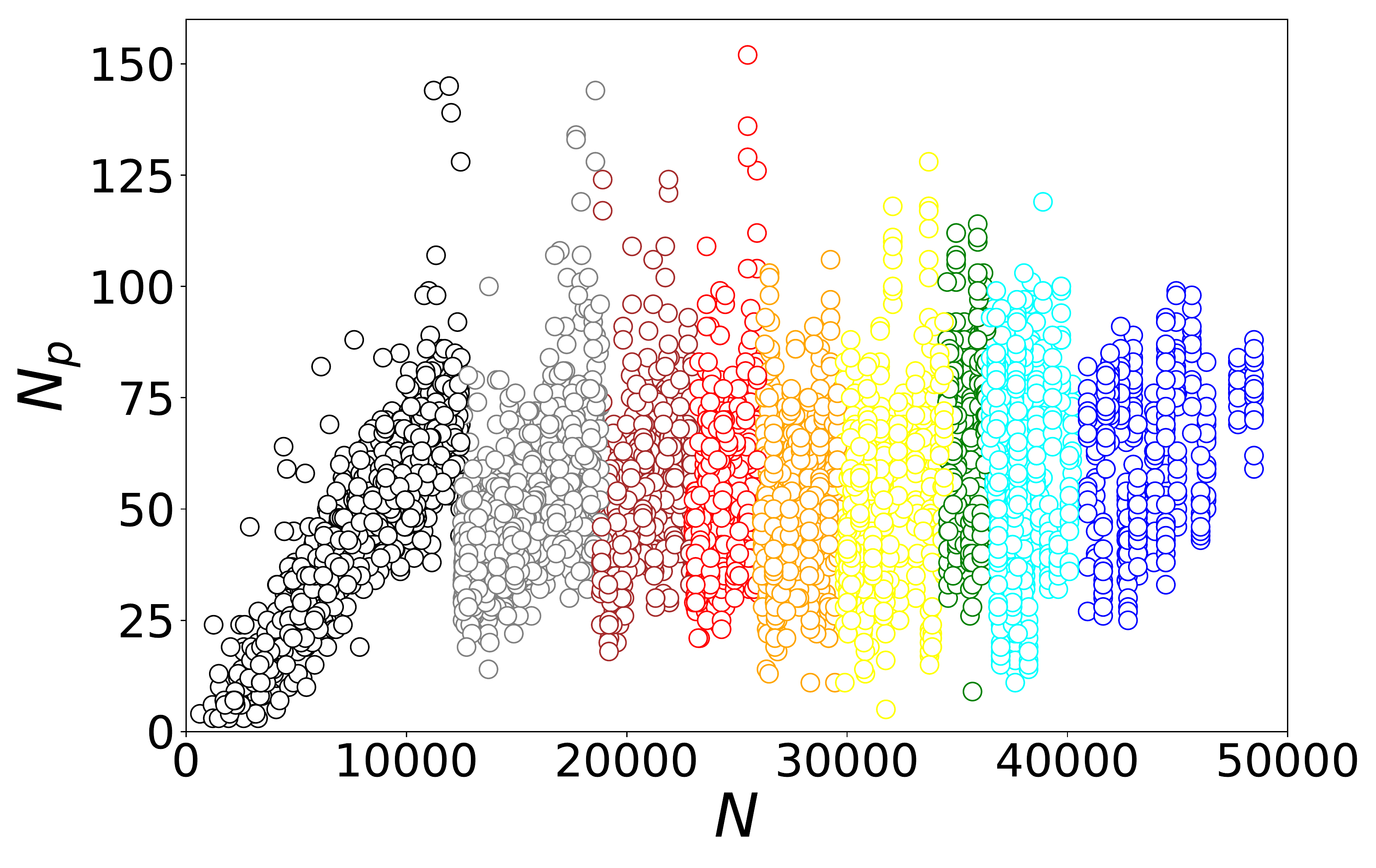}
\quad
\includegraphics[width=\columnwidth]{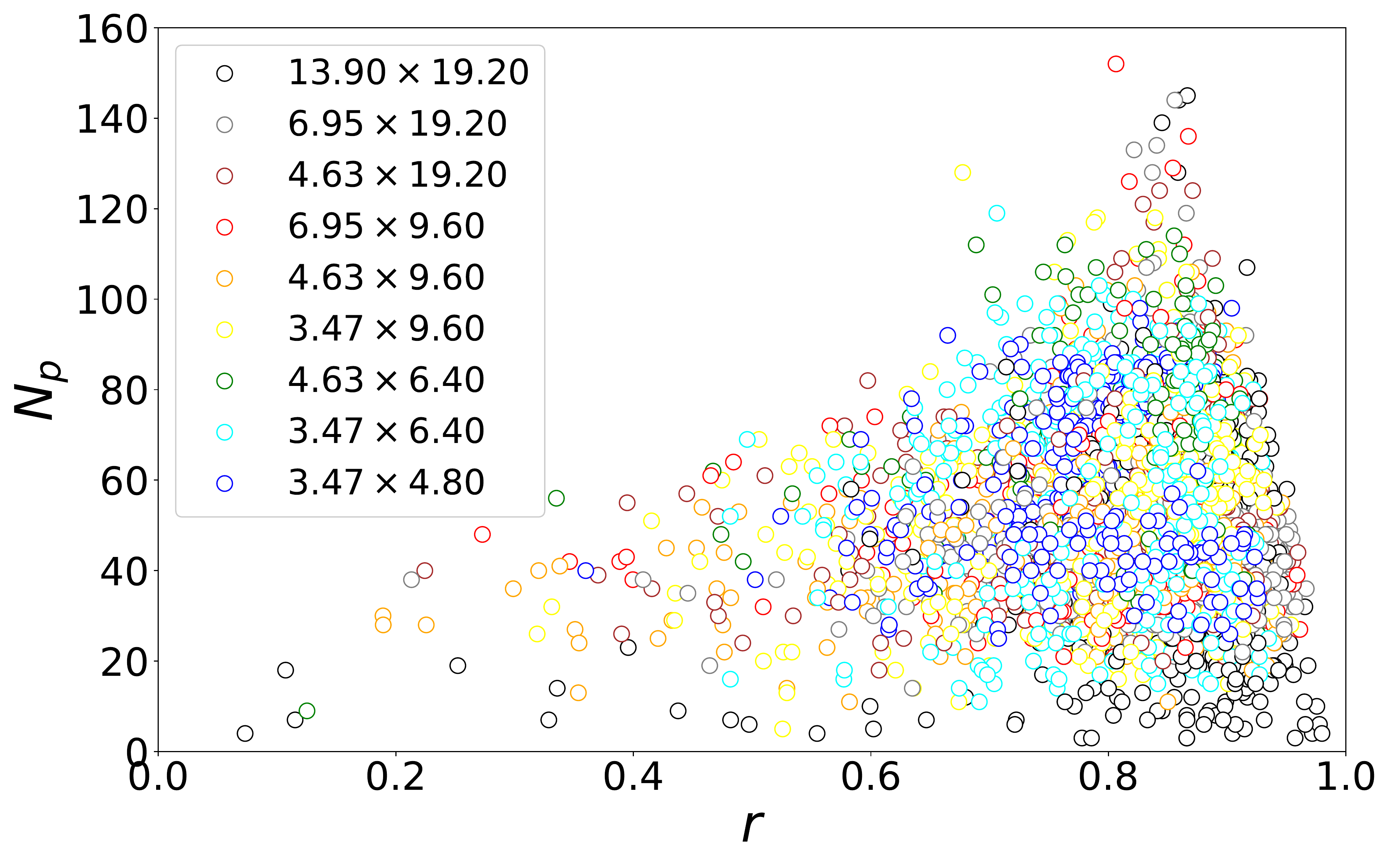}
\caption {(Top): $N_p$ versus $N$ for LMC subregions following finer
  binning. The colours correspond to the different bin areas; see
  Table \ref{table:tab1}, column (8). (Bottom): corresponding $N_p$
  versus $r$ distribution.}
\label{fig:div}
\end{figure}


Table~\ref{table:tab1} shows the nine criteria we adopted to divide
our LMC tiles so as to have similar ranges of $N_p$ (see
Fig.~\ref{fig:div}, top) for all criteria. Inspection of subregions
with high stellar density suggested that small-scale variations in
reddening and/or multiple dominant populations may cause the RGB to
broaden, resulting in poor $|$slope$|$ estimation. For regions with
lower stellar density (generally the outer regions) $N_p$ could be
low, leading to poorly defined RGBs and, hence, uncertain $r$. We thus
defined 4940 subregions, with areas ranging from
 13.90$\times$19.20~arcmin$^2$ (201.79$\times$278.73 pc$^2$) to
3.47$\times$4.80~arcmin$^2$ (50.37$\times$69.68~pc$^2$). This subdivision
is necessary (see C16), since very large $N_p$ ($> 100$) or small
$N_p$ ($< 20$) leads to poor value estimations of $|$slope$|$ with $r
< 0.5$. Fig.~\ref{fig:div} (bottom) shows the resulting $N_p$ versus
$r$ distribution.

\begin{figure}
\centering
\includegraphics[width=\columnwidth]{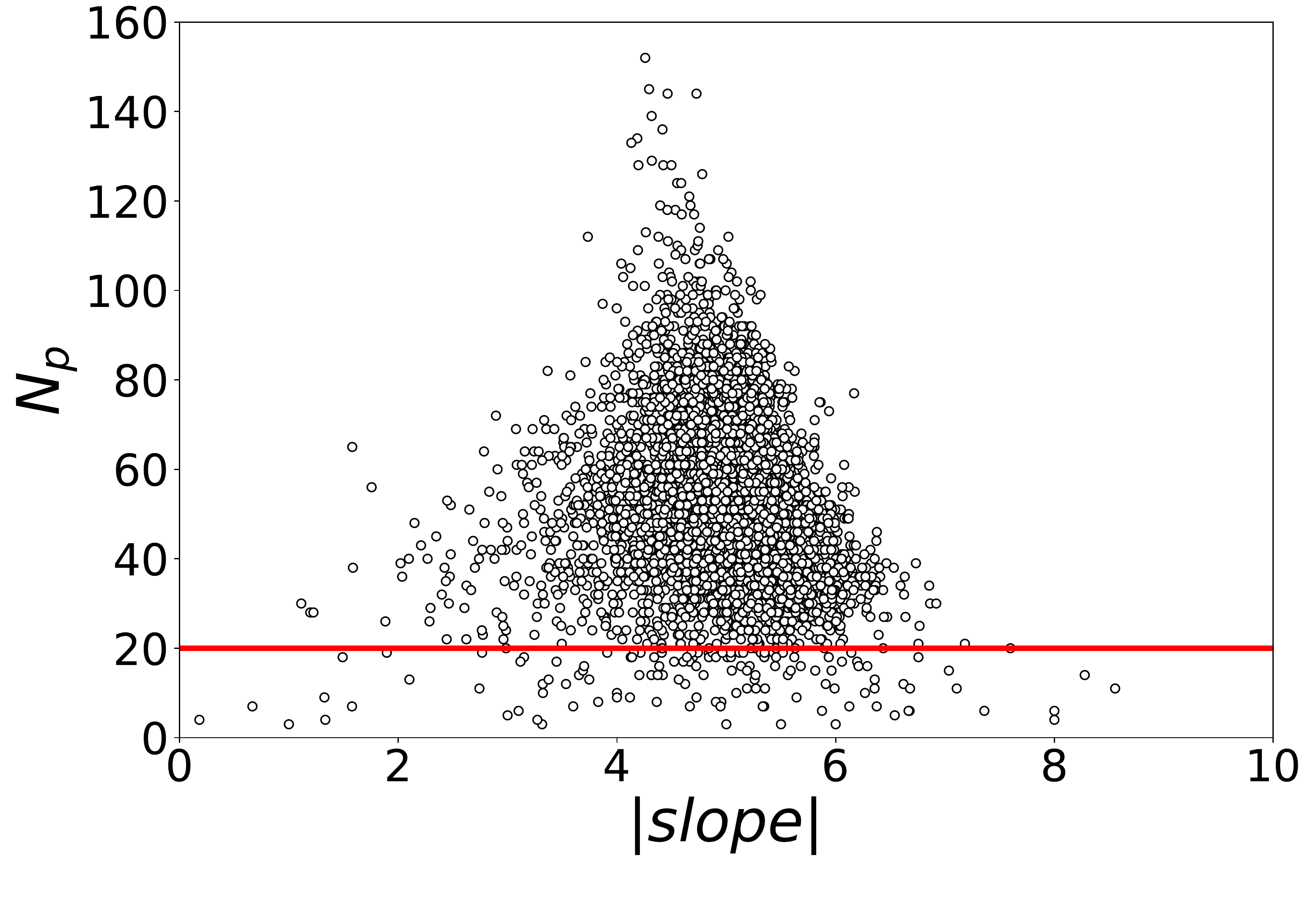}
\quad
\includegraphics[width=\columnwidth]{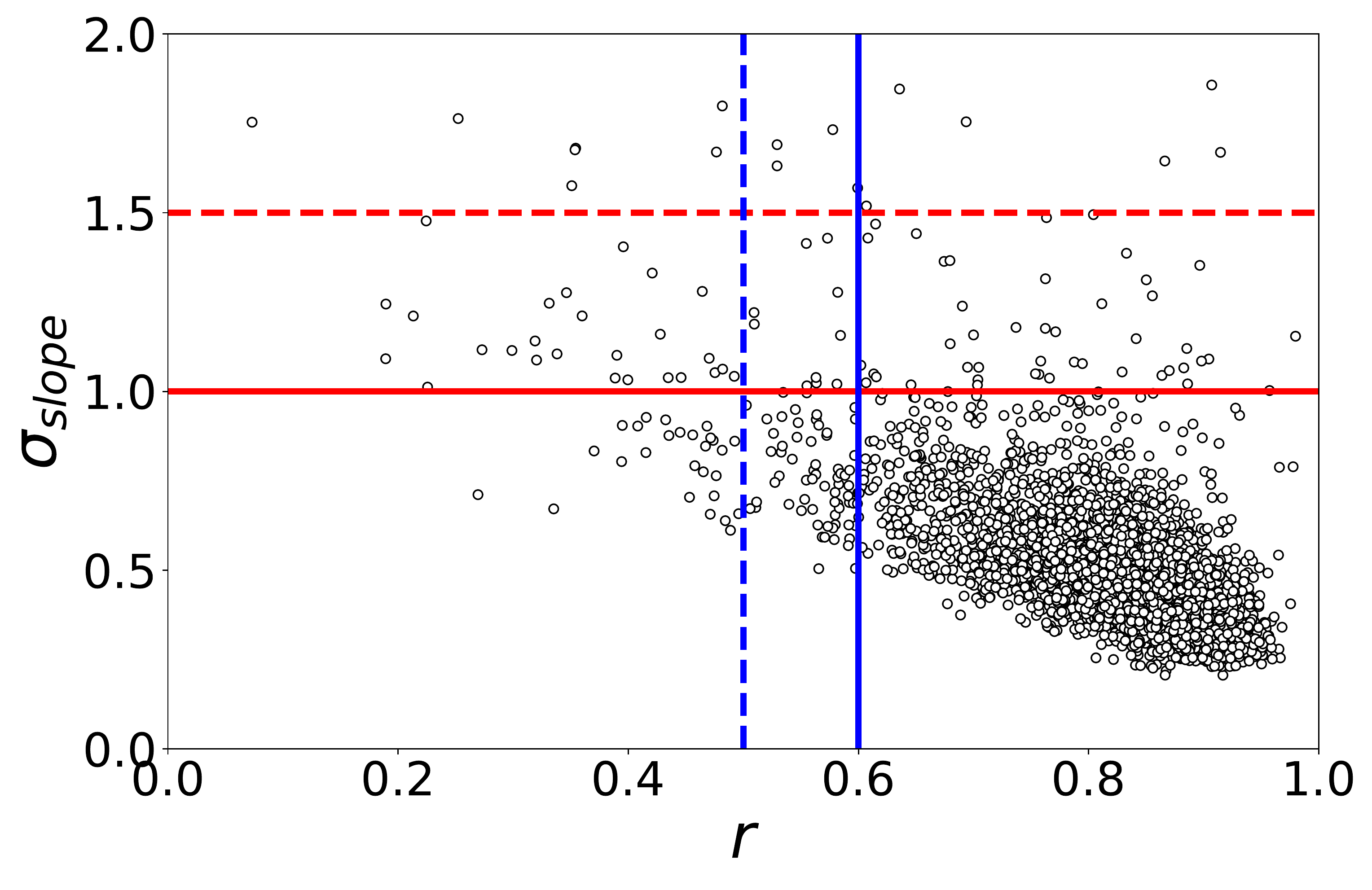}
\caption{(Top): $N_p$ versus $|$slope$|$ for the subregions. The red
  line at $N_p = 20$ denotes the cut-off adopted to exclude regions
  with poorly populated RGBs. (Bottom): $\sigma_{\rm slope}$ versus
  $r$ for the subregions. The red dashed and solid lines correspond to
  cut-off criteria for $\sigma_{\rm slope}$ of 1.5 and 1.0,
  respectively. The blue dashed and solid lines denote cut-offs
  corresponding to $r$ = 0.5 and 0.6, respectively.}
\label{fig:allcuts}  
\end{figure}

\begin{figure}
\includegraphics[width=\columnwidth]{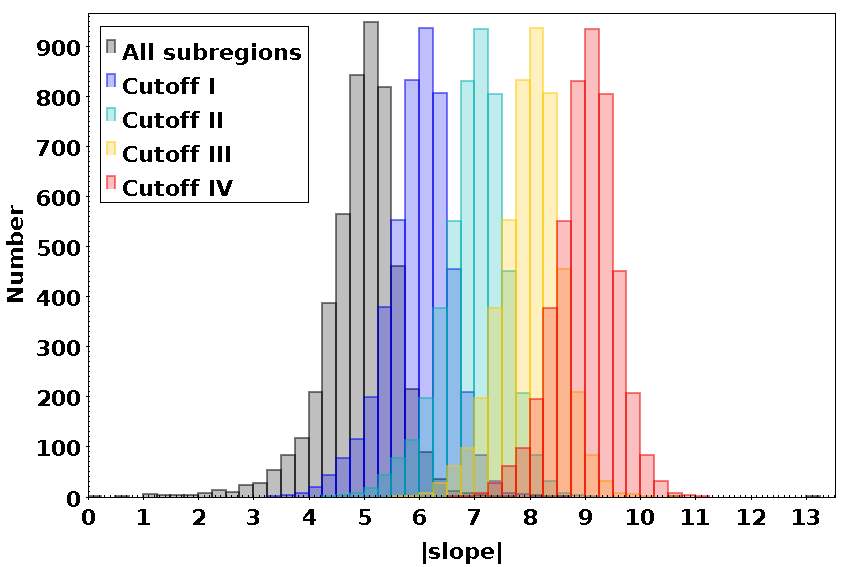}
\caption{$|$slope$|$ histogram for LMC subregions pertaining to all
  four cut-off criteria, compared with the equivalent distribution for
  no cut-off. Each histogram is shifted by 1.0 along the $x$ axis
    with respect to the previous histogram to avoid overlaps.}
\label{fig:hist_slope_allcuts}
\end{figure}

We sample against subregions with poor slope estimations by defining
four different cut-off criteria in terms of $r$, $\sigma_{\rm slope}$
and $N_p$. Fig.~\ref{fig:allcuts} (top) shows a large scatter in the
slopes for $N_p < 20$, which could be artefacts owing to sparsely
populated RGBs. In Fig.~\ref{fig:allcuts} (bottom), we note that the
clumpiest part of the distribution is found for $r > 0.6$ and
$\sigma_{\rm slope} < 1.0$. Most of the $r$ values are $>$ 0.5 for
$\sigma_{\rm slope} < 1.5$. We observe a relatively large scatter for
subregions with $r < 0.5$ and $\sigma_{\rm slope} > 1.5$. Thus, our
four cut-off criteria ($N_p \geq 20$ for all) are:
\begin{itemize}
\item criterion I: $r \ge 0.5$ and $\sigma_{\rm slope} \le 1.5$;
\item criterion II: $r \ge 0.5$ and $\sigma_{\rm slope} \le 1.0$;
\item criterion III: $r \ge 0.6$ and $\sigma_{\rm slope} \le 1.5$;
\item criterion IV: $r \ge 0.6$ and $\sigma_{\rm slope} \le 1.0$.
\end{itemize}
Fig.~\ref{fig:hist_slope_allcuts} shows a comparison of the RGB slope
distribution for all four cutoff criteria with respect to the original
sample. The distribution is curtailed on the side with lower RGB
slopes as we tighten our criteria. The peaks and overall shape of the
distribution are not affected significantly as we move from criterion
(I) to (IV). The numbers of subregions removed are 3.7 and 5.6 per
cent for criteria (I) and (IV), respectively. Thus, even with the
strictest criterion we end up with a statistically significant number
of subregions (4663) to proceed with our analysis. We point out
  that calibrating the PSF catalogue using photometric zero-points
  contained in v1.5 \citep{Gonzalez+2018MNRAS} instead of v1.3 will
  not affect our results, since any differences in the estimated RGB
  slopes will be well within the 1$\sigma$ uncertainties.

\begin{table*}
\caption{ The nine binning criteria used to subdivide LMC tiles.
  For each criterion, column (2) indicates the limit on the total
  number of stars ($N$) within a region. Column (3) lists the number
  of regions within that limit. Columns (4) and (5) specify the
  numbers by which regions are binned along RA and Dec,
  respectively. Column (6) lists the total number of
  subregions. Column (7) gives the area of each such subregion, and
  column (8) denotes the total number of subregions corresponding to
  each of the six subdivision criteria. The colours adjacent to the
  numbers are used to denote them in Fig.~\ref{fig:div} (top and
  bottom).}
\label{table:tab1}
\begin{tabular}{|c|c|c|c|c|c|c|c|}
\hline \hline
        & Stars   & Regions  & RA         & Dec        & No. of            & Area         &  Subregions    \\
        &         &          & divisions  & divisions  & divisions         & (arcmin$^2$) &  ($a \times d$) \\
        &         & (a)      & (b)        & (c)        & $(d= b \times c)$ &              &                 \\
\hline\hline
1  & 0 $<$ $N$ $\le$ 12,500        & 902 & 1 & 1 & 1  & (13.90$\times$19.20) & 902 (black)\\
2  & 12,500 $<$ $N$ $\le$ 18,800   & 538 & 2 & 1 & 2  & (6.95$\times$19.20) & 1076 (gray) \\
3  & 18,800 $<$ $N$ $\le$ 23,000   & 172 & 3 & 1 & 3  & (4.63$\times$19.20) & 516 (brown) \\
4  & 23,000 $<$ $N$ $\le$ 26,000   & 99  & 2 & 2 & 4  & (6.95$\times$9.60) & 396 (red)    \\
5  & 26,000 $<$ $N$ $\le$ 29,800   & 84  & 3 & 2 & 6  & (4.63$\times$9.60) & 504 (orange) \\
6  & 29,800 $<$ $N$ $\le$ 34,500   & 65  & 4 & 2 & 8  & (3.47$\times$9.60) & 520 (yellow) \\
7  & 34,500 $<$ $N$ $\le$ 36,500   & 26  & 3 & 3 & 9  & (4.63$\times$6.40) & 234 (green)  \\
8  & 36,500 $<$ $N$ $\le$ 40,500   & 34  & 4 & 3 & 12 & (3.47$\times$6.40) & 408 (cyan)   \\
9  & $N$ $>$ 40,500                  & 24  & 4 & 4 & 16 & (3.47$\times$4.80) & 384 (blue)   \\
\hline     
\end{tabular}
\label{table:tab1}
\vskip 1.0ex
\end{table*} 
%

\subsection{Calibration of the RGB slope as a function of metallicity}

We use the same spectroscopic studies as C16 for calibration: metallicities of RGs in the bar of the LMC
\citep{Cole+2005AJspectroOfRGs}, star clusters
\citep{Grocholski+2006AJCaIItriplet} and fields covering areas
  $6.^{'}8 \times 6^{'}.8 $ around these clusters (A. A. Cole, private
  communication). All three studies were conducted by the same group
  using the FORS2 instrument on ESO's 8.2~m Very large Telescope. The
  authors used the CaT lines to estimate the metallicities of their
  RGs, including the calibration of CaT strength to [Fe/H]. Thus,
  there are no inconsistencies or systematic offsets among these
  studies. Together, these three studies cover a range in metallicity
and location across the LMC, which is important for our calibration of
the RGB slopes. We use a similar technique as C16 (see their Section~3.1) to estimate the mean metallicities of the subregions and star
clusters/fields around them.

\begin{figure} 
\includegraphics[height=0.65\columnwidth,width=\columnwidth]{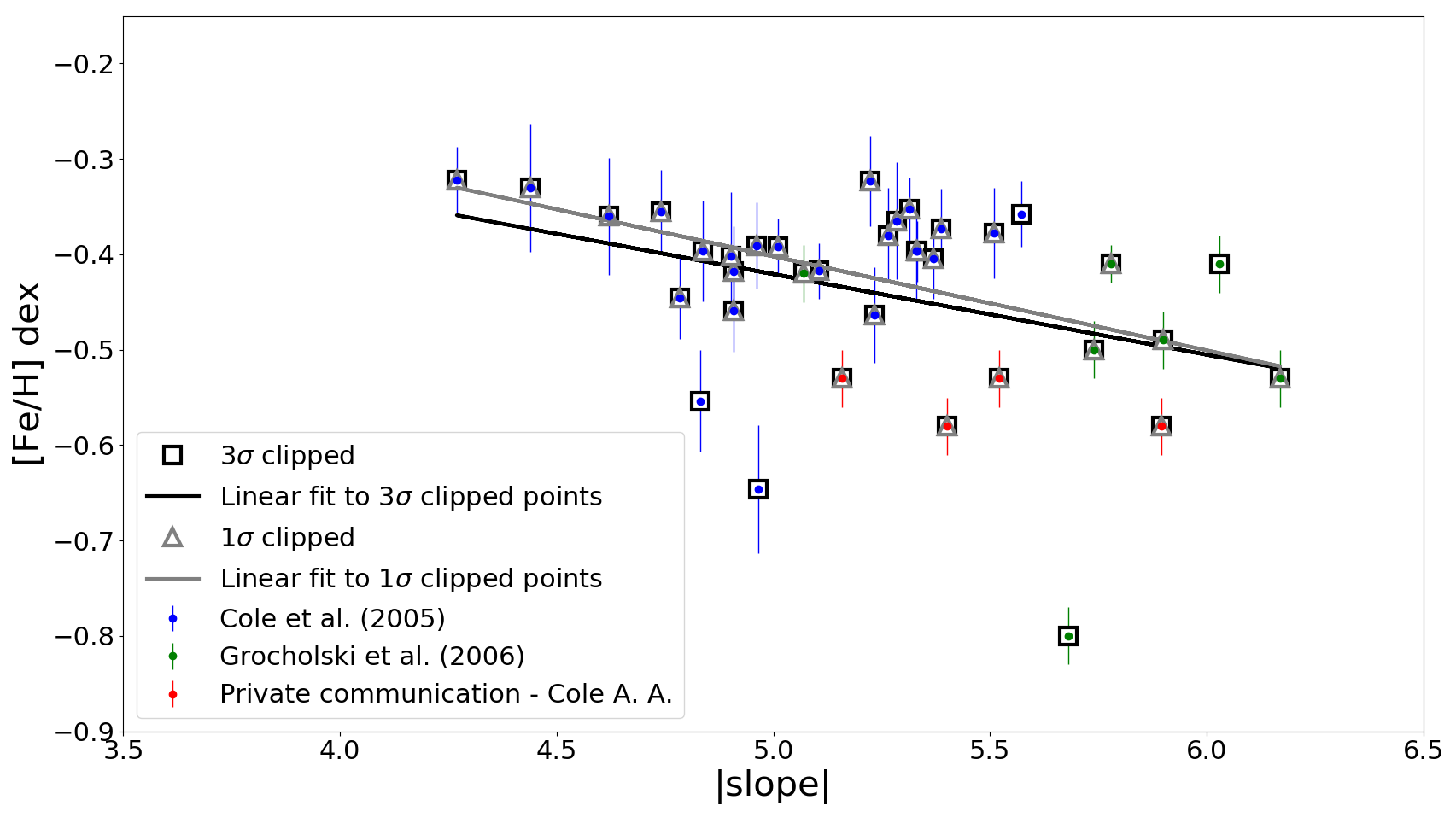}
\caption{Metallicity ([Fe/H]) versus $|$slope$|$. Blue points denote
  our subregions whose mean [Fe/H] has been found using RGs from
  \protect\cite{Cole+2005AJspectroOfRGs}, green points denote clusters
  from \protect\cite{Grocholski+2006AJCaIItriplet} and red points
  correspond to fields around
  \protect\cite{Grocholski+2006AJCaIItriplet}'s clusters provided by
  Andrew A. Cole. The grey open squares and solid line denote the
  3$\sigma$-clipped points and their corresponding best fits. The
  black open triangles and solid line denote the 1$\sigma$-clipped
  points and their corresponding best fit. The error bar
    (vertical coloured line) shown for each point is the standard
  error in the mean [Fe/H].}
\label{fig:lmccalib}
\end{figure}

\cite{Cole+2005AJspectroOfRGs} estimated the metallicities of 373
field RGs within a 200 arcmin$^2$ area at the optical centre of the
LMC bar. Their metallicity distribution is sharply peaked at the
median value, [Fe/H] = $-0.40$ dex, with a small tail of stars
extending down to [Fe/H] = $-1.21$ dex. We estimate the mean
metallicity for a subregion by averaging over the Cole et
al. metallicities within the relevant area. While doing so we
consider stars located within twice the standard deviation about the
mean metallicity. To ensure a good calibration, we consider those
subregions with $r \geq 0.70$, $\sigma_{\rm slope} \leq 0.5$ and which
contain spectroscopic metallicity estimates for at least five
RGs. These are indicated by blue points in the [Fe/H] versus
$|$slope$|$ plane shown in Fig.~\ref{fig:lmccalib}.

\cite{Grocholski+2006AJCaIItriplet} estimated the abundances of 28 LMC
clusters, aged between 1 to 13~Gyr, whose metallicities ranged between $-$0.30 and $-$2.0~dex. We
use the central coordinates and radii of those lying within the VMC
coverage to extract their data and construct $(Y, Y-K_{\rm s})$
  CMDs. We then employ our technique to estimate the RGB slopes for
these clusters. For most of the clusters, the RGB is either sparsely
populated or exhibits significant scatter about the mean RGB (either
due to crowding or mixing of cluster and field stars in the VMC data
or, less likely, differential reddening). For sparsely populated
clusters we relax our RGB identification criteria. We select
colour--magnitude bins with $N_p$ $\geq$ 2 stars (instead of 3) to
recover their RGB slopes. The RGB slopes and mean [Fe/H] values
  of the star clusters are shown as green points in
Fig.~\ref{fig:lmccalib}. A similar analysis is carried out for the
field stars, starting with their extraction from VMC fields from which
the cluster area is subtracted, estimating the RGB slopes from
$(Y, Y-K_{\rm s})$ CMDs. For some of the sparse fields, we identify
the RGB using a lower cut-off in colour--magnitude bins (i.e., $N_p
\geq$ 2 stars). The resulting RGB slopes and the mean [Fe/H]
  values of these fields are shown in red in Fig.~\ref{fig:lmccalib}.

In Fig.~\ref{fig:lmccalib} we see that the RGB slopes range from 4.25
to 6.25, covering most of the RGB slope
  distribution shown in Fig.~\ref{fig:hist_slope_allcuts}. The 
  mean metallicity range covers ranges from $-0.30$ to $-0.80$~dex. To estimate a slope--metallicity relation, we perform linear
least-squares fits with two different clipping choices, 3$\sigma$ and
1$\sigma$:
\begin{equation} \label{eq:1} 
{\rm [Fe/H]}= (-0.08\pm0.04)\times|{\rm slope}|+ (0.001 \pm 0.19)~\textrm{dex}, r=0.37;
\end{equation}
\begin{equation} \label{eq:2} 
{\rm [Fe/H]}= (-0.10\pm0.02)\times|{\rm slope}|+ (0.09 \pm 0.13)~\textrm{dex}, r=0.59.
\end{equation}
Their slopes and $y$-intercept values agree within the errors. The
1$\sigma$-clipped relation has a higher $r$ but a lower reliability of
the estimated slope--metallicity relation. We use both to estimate the
metallicities and compare our results in Sections~4 and 5. Our
  calibration of the RGB slope to metallicity rests on the assumption
that the spectroscopic targets are drawn from the dominant population
in the subregions. A direct comparison of the slope--metallicity
relation with C16 is not possible since the relation is a function of
the wavelengths used for these studies. We observe a scatter in the
plot resulting in lower $r$ compared with that of C16 where $r \sim
0.50$ for the 3$\sigma$-clipped case. However, for this study we used
twice more calibration points (36) compared with C16 (16). The
  total number of spectroscopic RGs used to estimate the
  3$\sigma$-clipped slope--metallicity relation is 486. We also
  investigated APOGEE spectroscopic data for our calibration, but the calibrators had insufficient range in RGB slope and metallicity values, to estimate a relation.

\begin{figure*} 
\includegraphics[height=3.5in,width=5.5in]{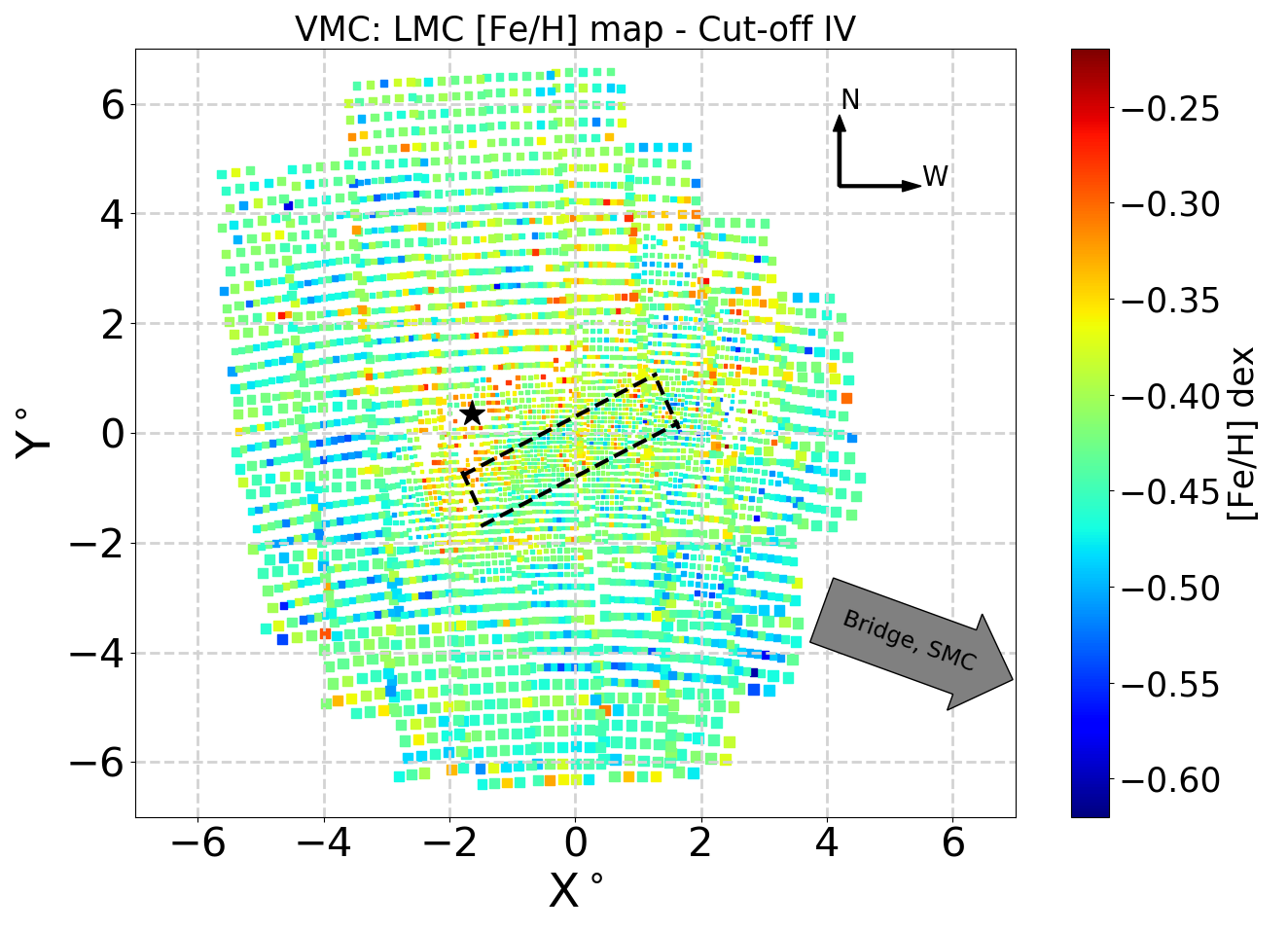}
\caption{Metallicity map based on cut-off criterion (IV) using
    Equation~(1). The bar region is shown by a dashed rectangle near
    the centre. The location of 30~Dor is shown by a black star.}
\label{fig:metalmap}
\end{figure*}
\quad
\begin{figure*}
\includegraphics[height=3.5in,width=5.5in]{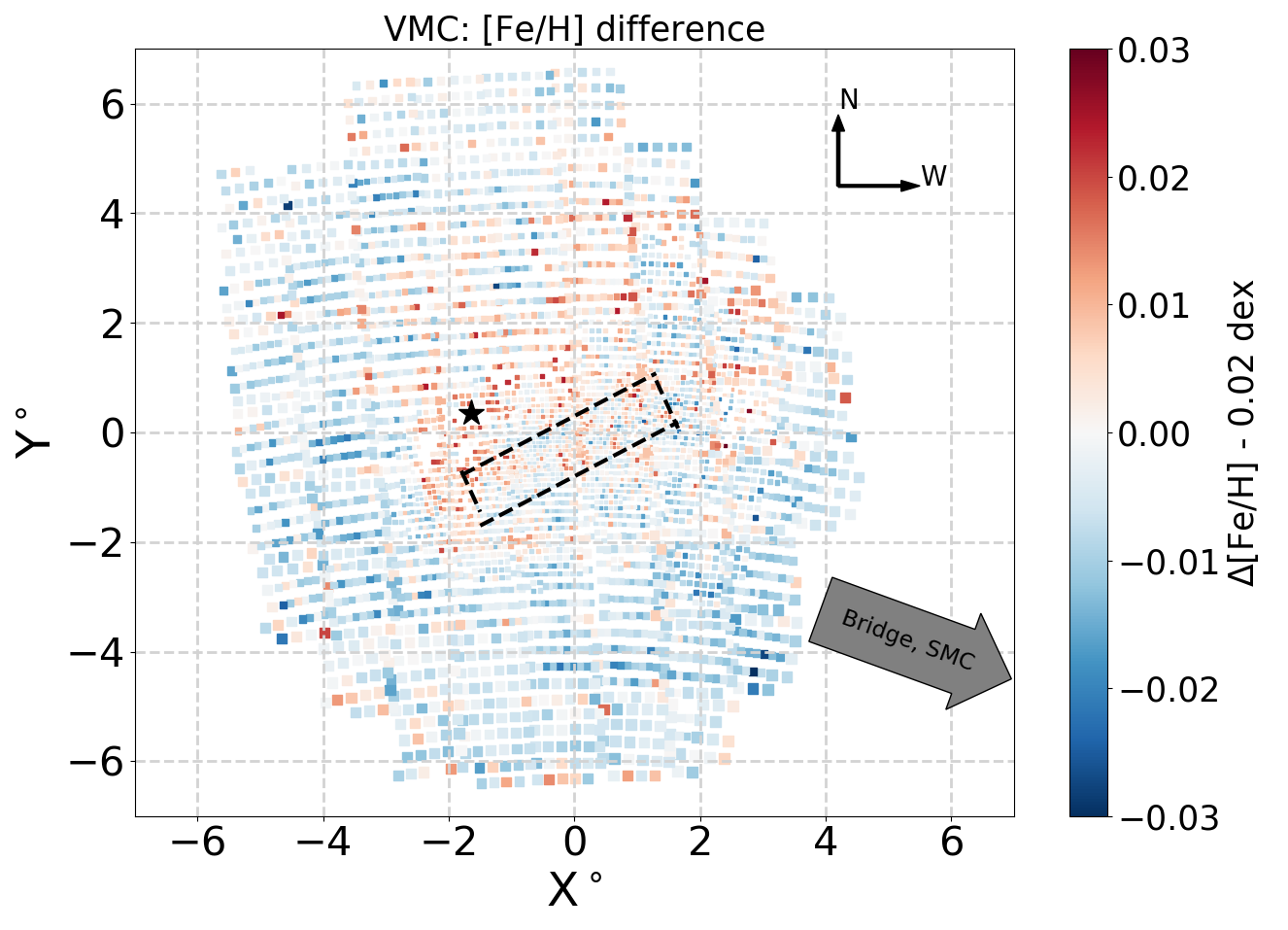}
\caption{Difference in metallicities, $\Delta$[Fe/H]. The
    difference is defined as `metallicities estimated using
    Equation~(1) subtracted from the corresponding values estimated
    using Equation~(2)'. The $\Delta$[Fe/H] values are shifted by
    $-$0.02~dex to bring the zero values to the centre of the colour
    bar. The location of the bar and 30~Dor are shown as in Figure
    \ref{fig:metalmap}.}
\label{fig:diffmetalmap}
\end{figure*}
\section{NIR metallicity maps of the LMC}

Fig.~\ref{fig:metalmap} shows the estimated metallicties for the LMC
subregions in the projected sky plane using Equation~(1). For the
LMC's centre we adopted RA = 5$^{\rm h}$ 19$^{\rm m}$ 38$^{\rm s}$ and
Dec = $-$69$^{\circ}$ 27$'$ 5.2$''$
\citep[J2000.0][]{deVaucouleurs&Freeman1972VAstructure}. The bar
  region shown in the figure is in accordance with
  \cite{Smitha&Purni2010A&Aanestimate}. We estimated metallicity maps for all
four cut-off criteria mentioned in Section~3. As we move from cut-off
criterion (I) to (IV), we lose 3.7 to 5.6 per cent of subregions. For
C16 the loss of subregions was 16.9 and 33.7 per cent for OGLE~III and
MCPS maps, respectively. Thus, we have a more or less uniform spatial
distribution with smaller gaps compared with the C16 maps. We
did not find any differences in global features among the four
NIR maps. Hence, we present the map with the strictest cut-off
criteria. The NIR metallicity map of the LMC covers an area that is
about thrice larger compared with the optical maps of C16. The
OGLE~III maps covered primarily the bar region and the eastern
  and western parts of the LMC out to $\pm$~4.0$^\circ$ along their
$x$ axes (see fig.~13 of C16). MCPS maps covered $\pm$~3.5--4.0$^\circ$ uniformly along the $x$ and $y$ axes (see fig.~29 of
C16). The VMC maps reveal trends out to $\pm$~6$^\circ$ along the $y$
axes and beyond and to $\pm$~4.5$^\circ$ along the $x$ axes.

The bar region is almost homogeneous with the more metal-rich points ([Fe/H]
$> -0.35$~dex) located near the 30~Doradus (30~Dor) star-forming region, and some
in the northwestern (NW) part of the bar. The most metal-poor regions
($<-0.55$~dex) are located in the outer disc (at radii
$>$2$^\circ$). We note a shallow variation of [Fe/H] from the bar to
the outer regions as compared with the maps of C16. The metallicity in
the NIR maps ranges from $-$0.22~dex to $-$0.60~dex, whereas the C16
maps showed a wider variation from $-$0.10~dex to $-$0.90~dex. We note
that the northern disc of the LMC has a larger range in metallicity
compared with the southern part. This is similar to what was observed
in the MCPS maps of C16. The eastern and western disc of the LMC
  have a similar range in metallicity, a trend which is also observed
  in the OGLE~III maps of C16. Some gaps appear in the maps; the most
  prominent ones correspond to the 30~Dor region and to regions North
of the bar. These regions most likely suffer from issues caused by
crowding or large variations in reddening. Such gaps are more
prominent in the optical maps of C16, and in particular for the MCPS
data set compared with OGLE~III. This is most likely related to poor
seeing conditions and/or the low resolution of the MCPS
images. However, with the NIR maps we have been able to cover most of
these gaps.

We also used Equation~(2) to estimate metallicities.
Fig.~\ref{fig:diffmetalmap} shows the map of differences between the
metallicities estimated using the two different slope--metallicity
calibration relations. For all subregions, Equation~(2) results in
enhanced metallicities compared with Equation~(1). This is owing to
the fact that Equation~(2) has a steeper slope and a more metal-rich
$y$ intercept compared with Equation~(1). The mean difference between
the metallicity estimates from these two calibration relations is
small, however, 0.018$\pm$0.007 dex. A byproduct of C16's work
  was a metallicity map of outliers, which identified significantly
  different metallicities in the LMC bar and outskirts. The metal-poor
  outliers were found in the northern, eastern and western disc,
  whereas the metal-rich counterparts were primarily seen in the bar
  region and in a few locations in the north. We were unable to
  estimate a map of metallicity outliers as in C16, since we recover a
  smaller metallicity range in this study. We checked our results by
plotting subregions that deviate by $\pm$1$\sigma$ about the mean
metallicity ($\sim$ $-$0.42~dex) of the LMC disc. This showed that the
most metal-rich points ($> -0.31$ dex) are located in the bar,
with only a few metal-poor ($> -0.53$~dex) points scattered in the
outskirts. CaT studies of 1000 LMC field giants by \cite{Olsen+2011ApJapopulation} found that a fraction of them represented a metal-poor, kinematically distinct population in the LMC's disc - possibly accreted from the SMC. However, like C16, we did not find any correlation between the subregions identified as metal-poor outliers and the locations in
which \cite{Olsen+2011ApJapopulation} discovered a possible accreted SMC population. This is possibly because at none of these
locations the kinematically distinct population is dominant enough to
turn the mean metallicity of the subregions significantly metal-poor
relative to neighbouring fields. This could be because the total
accreted population is either very small or well-mixed into the LMC
disc as a whole.

We also compare our metallicity maps with the most recent results of
\cite{Grady+2021ApJ}. Their LMC metallicity map extends out to
12$^\circ$ and shows a central metal-rich bar with the most metal-poor
regions littering the outskirts. The global features in their map are
generally consistent with those of C16 and the present study. Also, we
observe the northern and southern portions of the LMC disc to be
metal-enhanced compared with the eastern and western parts,
which is consistent with the results of \cite{Grady+2021ApJ} and
C16. Those authors also found metal-rich structures tracing the
  main spiral arm in the north and a prominent spiral-like feature in
  the southern portion of the disc. \cite{Dalal+2019MNRAS}'s
  morphology map of the LMC's RGB population (their fig.~6) shows an
  enhanced bar feature with some diffuse structures around the
  bar. For younger populations ($\lesssim 1$~Gyr old) the bar is less
  prominent and spiral features appear in the northern and southern
  regions of the disc. It is possible that our NIR metallicity maps
  represent a population with ages older than 1~Gyr.

\begin{figure}
\includegraphics[width=\columnwidth]{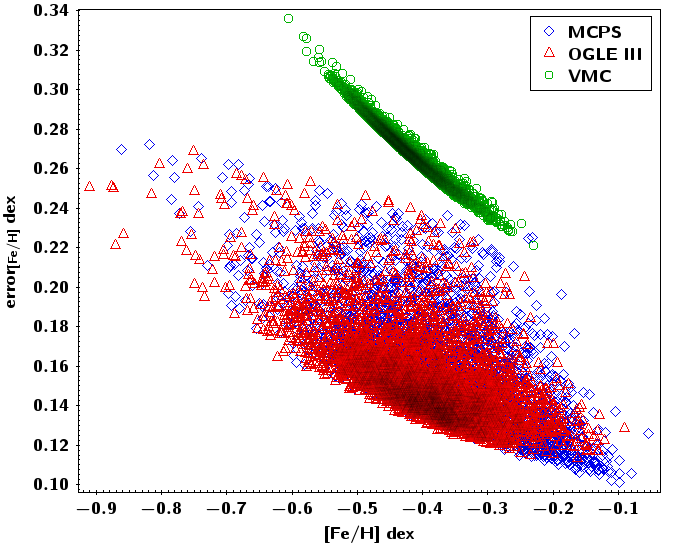}
\caption{${\rm error}_{\rm [Fe/H]}$ estimated in this study compared
  with C16 (OGLE~III and MCPS).}
\label{fig:Fe_H_errors}
\end{figure}
\begin{figure*} 
\includegraphics[height=0.25\linewidth]{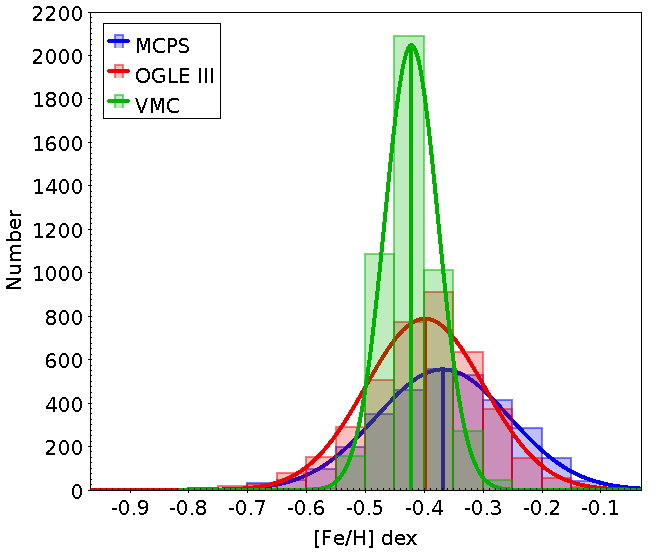}
\includegraphics[height=0.25\linewidth]{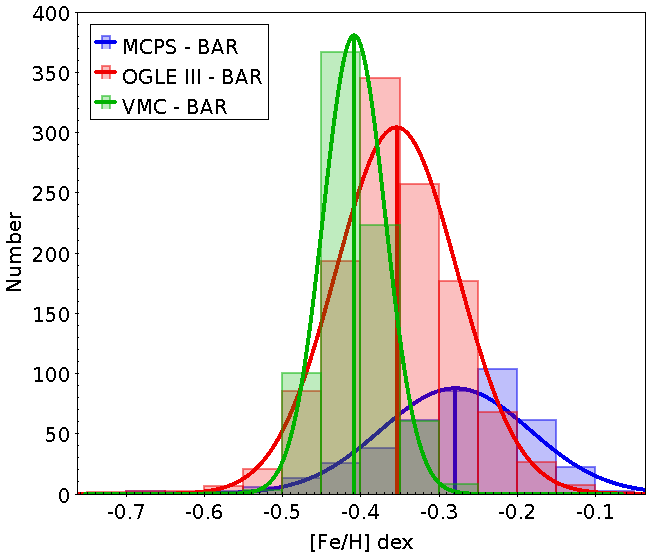}
\includegraphics[height=0.25\linewidth]{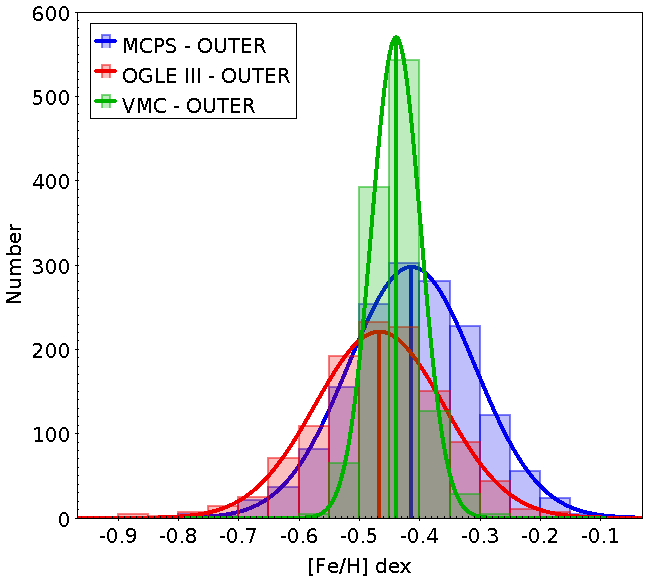}
\caption{Comparison of the metallicity distributions across the LMC
  between this VMC study and those of C16 (OGLE~III and MCPS) for
  different regions in the galaxy: left -- complete coverage of each
  survey; middle -- LMC bar; right -- outer disc (see text for
  definitions).}
\label{fig:metaldistribute}
\end{figure*}
We have calculated the errors associated with our metallicity
estimates (${\rm error}_{\rm [Fe/H]}$) using error propagation
following equation~(4) of C16 (see their
section~5). Fig.~\ref{fig:Fe_H_errors} shows that the range of ${\rm
  error}_{\rm [Fe/H]}$ spans from $\sim$0.22~dex to 0.34~dex, i.e.,
the errors are larger than those obtained by C16 for the OGLE~III and
MCPS data sets. This is most likely related to the difference in the
corresponding values of $|$slope$|$ and the slope--metallicity
calibration relation, which are functions of the wavelengths used in
these studies. We also note a slight trend suggesting that our
estimated errors are larger for metal-poor values. We find that the
dispersion in the value of ${\rm error}_{\rm [Fe/H]}$ for a given
value of [Fe/H] is smaller ($\sigma \sim 0.01$~dex) for VMC compared
with optical data ($\sigma \sim 0.02$--0.03~dex). Referring back to
the error propagation equation of C16, we point out that this
dispersion is determined by the range of $\sigma_{\rm slope}$
associated with the corresponding RGB $|$slope$|$.

\section{Discussion}
 
\subsection{Impact of reddening and distance variations}

Studies in the literature show that 30~Dor and the central region
  of the LMC have higher reddening compared with the rest of the disc
  (e.g., \citealt{Zaritsky+2004AJmcps,
    Haschke+2011,Smitha&Purni2013A&A, Gorski+2020ApJ, Skowron+2021ApJS,
    Grady+2021ApJ}). Recent extinction maps of \cite{Mazzi+2021} based
  on star-formation history (SFH) studies using VMC data also support these previous
  findings. The variation in reddening among subregions can shift the
  location of the RGB in the respective CMDs. Since we have adopted
  the technique of C16, this effect is accounted for in our analysis
  by anchoring the RGB to the densest part of the RC. However, large
  differential reddening variations within a subregion will broaden
  the RGB, leading to poorly estimated slopes. The effect of reddening
  is more significant in optical passbands. This is prominently
  observed in the metallicity maps of C16, where gaps are seen in the
  bar and in star forming regions like 30~Dor. NIR passbands offer an
  advantage since they suffer less from reddening compared with
  optical passbands. However, we still notice some gaps in our NIR
  metallicity maps, primarily in 30~Dor and near the bar region, which
  suggests that our technique fails to properly deal with subregions
  exhibiting large variations in reddening. Since the number of such
  subregions is between 3 and 6 per cent of our sample, their impact
  on our estimated mean metallicity and the radial metallicity
  gradient of the LMC is negligible.

  In their distance modulus map \cite{Mazzi+2021} show the distance
  variation among their adopted subregions in the LMC's disc. Our
  subregions are much smaller ($\sim$ 2 -- 28 times) compared with
  \cite{Mazzi+2021}. Thus, it is safe to assume that there is no
  significant distance variation within a given subregion. By virtue
  of anchoring the RGB to the densest part of the RC for a subregion,
  our technique accounts for distance variations among subregions.

\subsection{Metallicity distribution within the LMC}

In Fig.~\ref{fig:metaldistribute}, we show a comparison of the
metallicity distributions for the complete LMC, the bar region and an
outer region estimated from VMC data and with the results of C16. When
referring to the outer LMC, we mean the regions that lie beyond a
radial distance of 2.5$^\circ$ from the LMC centre. We estimated the
mean metallicity by fitting Gaussian profiles to all
distributions. Overall we see that the width of the VMC distribution
($\sigma[Fe/H]$= 0.04~dex) is much smaller than the MCPS and OGLE~III
distributions ($\sigma$[Fe/H]= 0.10--0.11~dex) for all three specified
regions in the LMC. The mean metallicity estimated using VMC data
([Fe/H]=$-$0.42~dex, $\sigma$[Fe/H]= 0.04~dex) is relatively
metal-poor compared with MCPS ([Fe/H]=$-$0.37~dex, $\sigma$[Fe/H]=
  0.11~dex) and OGLE~III data ([Fe/H]=$-$0.39~dex,
  $\sigma$[Fe/H]= 0.10~dex), which peak at almost similar values. The
range of metallicities estimated using VMC data is smaller compared
with those obtained using MCPS and OGLE~III data. Thus, we have used
finer binning (0.05~dex) to construct our histograms than C16
(0.15~dex) to compare with OGLE~III and MCPS results. However, note
that the mean metallicity and the widths of the OGLE~III and MCPS
distributions are not affected by this smaller binning.

The mean metallicities of the bar region as estimated from the MCPS,
OGLE~III and VMC data are $-$0.28~dex ($\sigma$[Fe/H]=
  0.09~dex), $-$0.35~dex ($\sigma$[Fe/H]= 0.08~dex) and $-$0.41~dex
($\sigma$[Fe/H]= 0.04~dex), respectively. The mean metallicity of the
outer LMC is $-$0.41~dex ($\sigma$[Fe/H]= 0.10 dex), $-$0.46~dex
($\sigma[Fe/H]$= 0.11~dex) and $-$0.44 dex ($\sigma$[Fe/H]= 0.04~dex),
for MCPS, OGLE~III and VMC, respectively. For the MCPS and OGLE~III data, the mean metallicity of the bar is prominently metal-rich as compared with the outer region. However, for the VMC data, the mean metallicities of the bar and outer regions are indistinguishable within the errors. The bar metallicity has lower $\sigma$[Fe/H] than the outer regions, which is detectable when
comparing the optical data sets. However, we cannot resolve this
difference in $\sigma$[Fe/H] for the VMC data set. 

Even though similar passbands ($V$ and $I$) were used for OGLE~III and
MCPS data, and the same calibration relation was used (after
correcting for systematic effects), C16 estimated smaller values of
MCPS slopes in the bar region, leading to higher values of
metallicity. The central regions are affected by crowding and/or
differential reddening. The MCPS data set is perhaps more affected by these
issues owing to its low resolution and/or poor seeing conditions. The
effect is clearly seen for common stars in the bar region in the
MCPS data compared with the OGLE~III and VMC data. We will discuss
this in more detail in the next section. A significant fraction of bar
subregions are missing in MCPS maps compared with OGLE~III, implying
that the bar region is poorly sampled by the MCPS. According to C16,
the MCPS maps are representative of the outer LMC owing to their
uniform coverage in all four directions, while the OGLE~III maps
represent the metallicity trend in the bar. The VMC maps provide a
more comprehensive picture of both the bar and the outer regions of
the LMC.

\subsection{A metallicity gradient within the LMC}

\begin{figure}
\includegraphics[width=\columnwidth]{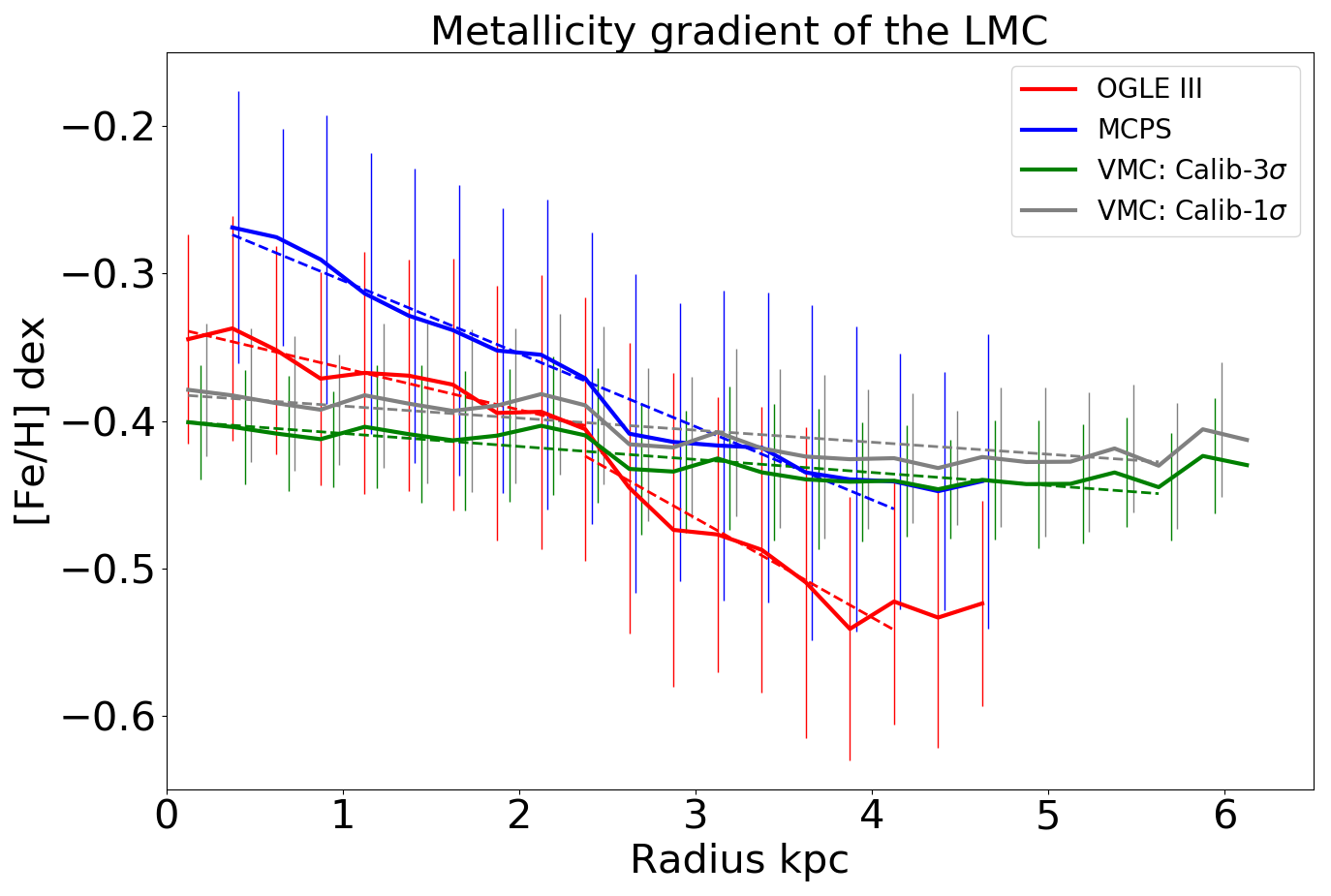}
\caption{Radial MG estimated in NIR passbands from the VMC survey
  shown with respect to estimations from optical passbands (OGLE~III,
  MCPS) from C16. The error bars denote the standard deviation
    about the mean. They are offset slightly along the $x$ axis for
    each case to avoid overlap.}
\label{fig:metgrad}
\end{figure}

To estimate the radial MG, we consider the LMC disc to be inclined
with respect to the sky plane by an angle $i$, and the position angle
of the line of nodes is given by $\Theta$. We use $i$ = $25^{\circ}.7$
and $\Theta$ = $141^{\circ}.5$ \citep{Smitha&Purni2013A&A}. We assume
the distance of the LMC to be 49.9~kpc \citep{deGrijs+2014AJ}. We use
these parameters in the equations of \cite{vander2001AJ-MCstructure},
to estimate the metallicity variation in the LMC plane. Then, we
estimate a radial MG by dividing the galaxy radially into bins of
0.25~kpc width. We also re-estimate the MGs resulting from the
OGLE~III and MCPS data sets (C16) out to a radius of about 4~kpc. To
avoid issues owing to poor sampling, we exclude the most distant
radial bins for all three data sets. In addition, for MCPS we avoid
the central radial bin to avoid small-number
statistics. Fig.~\ref{fig:metgrad} shows a comparison of the LMC MGs
for all three data sets. There is an indication of decreasing
  metallicity with increasing distance for the VMC data. However, the
variation is shallow compared with C16. Following C16, MG estimation
for OGLE~III is split into two parts, i.e. the central disc
  ($\lesssim$2.5~kpc, containing the bar) and the outer disc,
$-$0.028$\pm$0.003 and $-$0.057$\pm$0.009~dex kpc$^{-1}$,
respectively. The MCPS gradient is estimated at $-$0.049$\pm$0.002~dex
kpc$^{-1}$. The newly estimated MGs for OGLE~III and MCPS are very
similar to those of C16, although the geometric parameters assumed in
their studies ($i$ = $37^{\circ}.4$, $\Theta$ = $141^{\circ}.2$)
  were different. Using the VMC data we estimate a MG of
$-$0.008$\pm$0.001~dex kpc$^{-1}$ out to a radius of $\sim$6~kpc, with
a $y$ intercept of $-$0.40$\pm$0.004, and $r$= 0.91. The trend
estimated using the 1$\sigma$-clipped calibration relation
(Equation~2) gives MG = $-$0.010$\pm$0.001~dex kpc$^{-1}$. This
signifies that use of either of our different slope--metallicity
relations has a negligible effect on the estimated MG.

The mean metallicity per radial bin seems to be in agreement for the
three data sets owing to large standard deviation about the
mean. However, the differences are larger in the innermost (0 to
1~kpc) and outer ($>$ 4--5~kpc) parts of the galaxy. The mean
metallicity per radial bin is higher in the inner regions for MCPS,
whereas OGLE~III and VMC agree within the errors. At larger radii the
MCPS and VMC values are similar, but OGLE~III values are relatively
metal-poor. We inspected this to check whether this might be owing to
differences in the areas for the subregions adopted for this study and
in C16. This had a negligible effect on their studies, since they were
not interested in the average metallicity of each subregion, but
instead they determined the global average and its variation across
the LMC. Since the depth and resolution of MCPS and OGLE~III data
  are different, C16 used different area binning criteria for both
  data sets. Thus, at a given location the sizes of OGLE~III and MCPS
  subregions are different. C16 found that the difference between the
  RGB slope values (hence metallicity) at a given location for
  OGLE~III and MCPS were well within the mutual error bounds.

\begin{figure}
\includegraphics[width=\columnwidth]{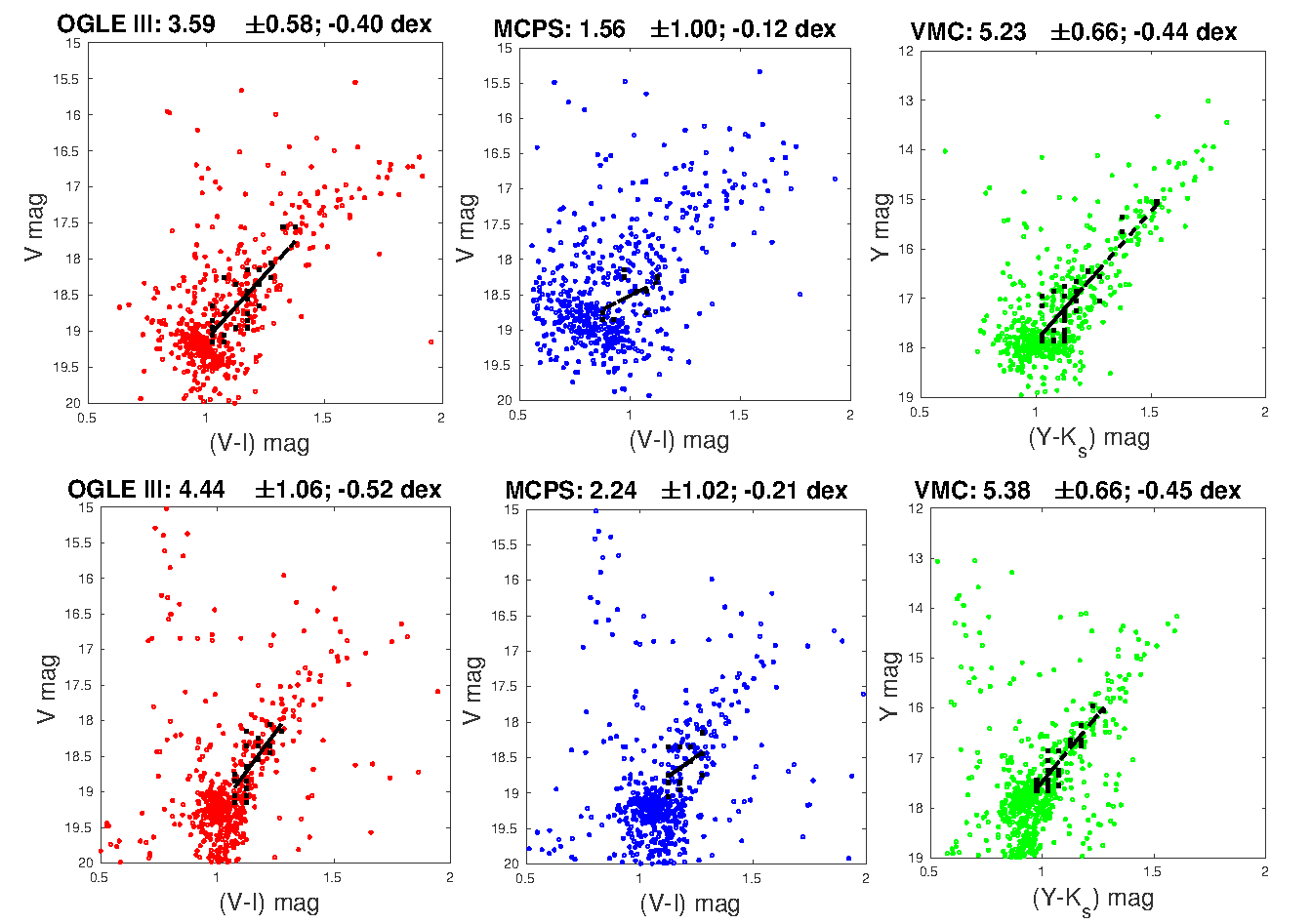}
\caption{CMDs of common stars for OGLE~III, MCPS and VMC data
  for a sample subregion, within the inner LMC (0~kpc to
  1~kpc, top) and the outer LMC (4~kpc to 5~kpc,
  bottom). The corresponding RGB slopes and [Fe/H] values are 
    listed in the panels' titles.}
\label{fig:xmatch_cmds}
\end{figure}

To check the effect of variable areas, we selected subregions from the
three data sets within the inner (0 to 1~kpc) and outer ($>$ 4--5~kpc)
LMC which had mean metallicities per radial bin similar to that
observed in Fig.\ref{fig:metgrad}. For subregions at radii from 0 to
1~kpc, we sampled subregions with metallicities between $-$0.250 to
$-$0.30~dex for MCPS; $-$0.325 to $-$0.375~dex for OGLE~III; and
$-$0.375 to $-$0.425~dex for VMC. For subregions at radii from 4 to
5~kpc we used $-$0.40 to $-$0.45 dex for MCPS; $-$0.50 to $-$0.55 for
OGLE~III; and $-$0.40 to $-$0.45~dex for VMC.  As an example,
  CMDs of a subregion in the inner and outer LMC are shown in
  Fig.~\ref{fig:xmatch_cmds} for the stars in common among the three
  data sets. We estimated the slopes for the OGLE~III and MCPS data
sets following C16 and calibrated them to metallicities using their
equation~(1). The VMC slopes were estimated using the technique
outlined in this study and calibrated by our Equation~(1). We find
that the differences in metallicity between the MCPS compared with
OGLE~III and VMC remain and are similar to what is observed in the
mean metallicity per radial bin in the inner LMC. The same is noted
between OGLE~III compared with MCPS and VMC in the outer LMC. The CMDs
also show that the MCPS data suffer from crowding issues and/or poor
seeing conditions. We checked this for additional subregions. This
effect is more pronounced in the central regions compared with
the outer regions, leading to a broadening of RGB and a shallow RGB
slope, hence metal-rich regions for MCPS. Thus, we suggest that at
optical wavelengths the OGLE~III results are more reliable than their
MCPS counterparts. Thus, the difference in trends observed in
Fig.~\ref{fig:metgrad} is not caused by differences in the areas
covered by the subregions. This is possibly related to the systematic
differences between the filter systems (optical and NIR), and
the slope--metallicity calibration relation, which is a function of
wavelength.

\subsubsection{Comparison with other studies}

The MG estimated by \cite{Cioni2009A&Athemetallicity} out to a
distance of 8~kpc using the C/M ratios of the field AGB population is
$-0.047\pm0.003$~dex kpc$^{-1}$, with an $y$ intercept of
$-1.04\pm0.01$ dex. Their radial MG is more consistent with OGLE~III
and MCPS than with our VMC results. This is perhaps related to
  the age difference ($\sim$ a few Gyr) between AGB and RGB stars. In
  their fig.~2, \cite{Cioni2009A&Athemetallicity} also show the
radial variation in spectroscopically derived metallicities of field
RGBs (from \citealt{Cole+2005AJspectroOfRGs, Pompeia+2008A&Achemi,
  Carrera+2008AJ-CEH-LMC}) and star clusters (from
\citealt{Grocholski+2006AJCaIItriplet,Grocholski+2007AJdist}). This
hinted at a negligible radial MG for RGBs within the inner 6~kpc of
the LMC, where the points are metal-rich compared with the AGB
distribution. A few points of the spectroscopic sample lying beyond
6~kpc were metal-poor and followed the MG trend of the AGB stars.
This trend of an apparently negligible MG out to 6~kpc agrees
  with our study. However, since the spectroscopic sample is
  statistically small compared with our photometric sample, we refrain
  from drawing significant conclusions from this similarity. Our
results are also shallow compared with \cite{Grady+2021ApJ}, who
estimated a MG of $-0.048\pm0.001$ dex kpc$^{-1}$, with a $y$
intercept of $-0.656\pm 0.004$ dex for LMC RGB stars out to a radius
of $\sim$12~kpc (their fig.~10). These authors used machine
  learning techniques to combine photometric ({\it Gaia} DR2, 2MASS,
  WISE) and spectroscopic estimates (APOGEE). While estimating the MG
these authors excluded the central regions (within $\sim$3~kpc) so as
to avoid the metal-rich bar and focus on the LMC disc. In doing so,
they mitigated the fact that their metallicity predictions for the
most metal-rich stars, which dominate centrally, incurred a degree of
bias ($\sim$0.1--0.2 dex) in their regression model.

\subsubsection{Asymmetric radial metallicity gradient}

\begin{figure}
\includegraphics[width=\columnwidth]{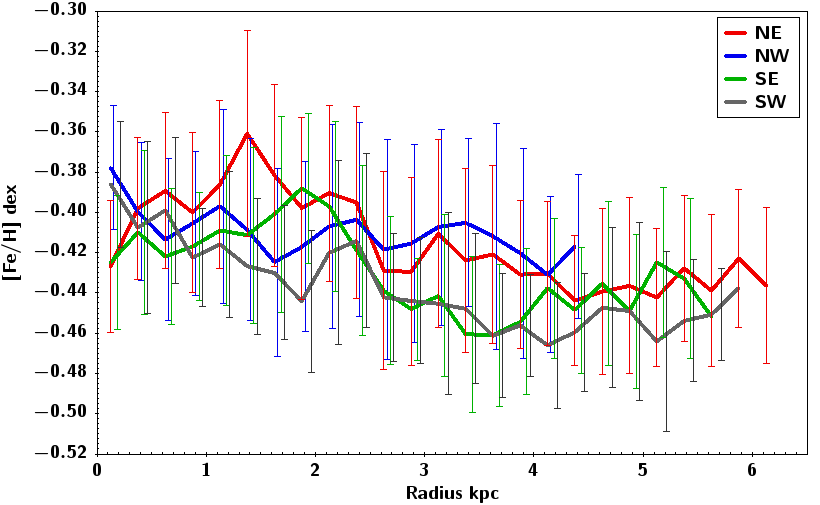}
\caption{Radial MGs in all four directions. The errors represent
    the standard deviation about the mean in each radial bin. The
    error bars have been shifted slightly along the $x$ axis to avoid
    overlap.}
\label{fig:PA_Metgrad}
\end{figure}

To check if there is any asymmetry in our radial MG, in
Fig.~\ref{fig:PA_Metgrad} we show the radial MG for four different
directions in the LMC with respect to the optical centre. The radial
trend is estimated out to 6~kpc for the North East (NE), South East
(SE) and SW quadrants, but for the NW region only out to 5~kpc. The
overall trends in all four quadrants are shallow and the mean
metallicity per radial bin agrees within the errors. However, their
appearances are suggestive. In the eastern quadrants we see a
metal-rich plateau followed by a dip in metallicity values beyond the
inner 2.5--3~kpc. The plateau for the NE quadrant is more metal-rich
compared with the SE and western quadrants. The trend in the SW
  portion, in the direction of the Bridge and the SMC, is relatively
metal-poor compared with all quadrants.

\begin{figure}
\includegraphics[width=\columnwidth]{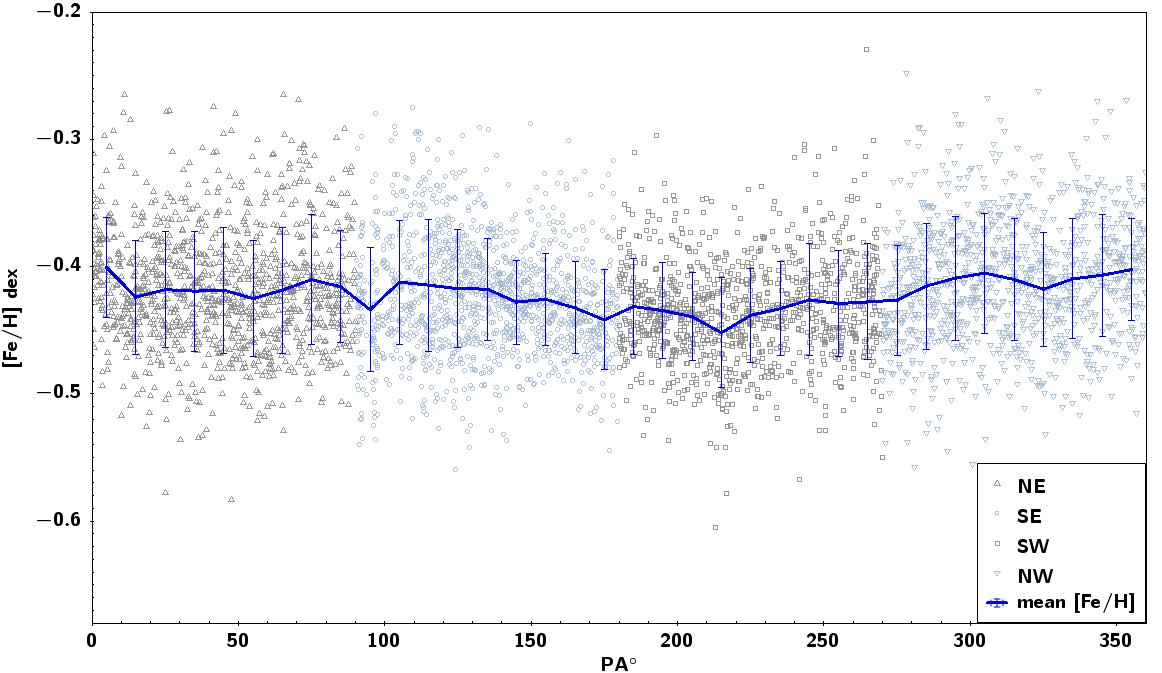}
\caption{Metallicity distribution as a function of PA for all four
  directions. The mean metallicity in all PA bins are shown as blue
  points. The standard deviation about the mean is shown as error
  bars. }
\label{fig:Metal_vs_PA}
\end{figure}

We also checked the variation in metallicity with respect to the
position angle (PA). Fig.~\ref{fig:Metal_vs_PA} shows that overall all
points are scattered between $-$0.6~dex to $-$0.2~dex. However, we
observe more metal-poor points in the southern LMC, especially in the
SW portion (180$^\circ$ to 270$^\circ$) compared with the northern
regions. We estimated mean metallicities by binning along the PA and overplotted them to check for any variations. Although these values for
each PA bin are in agreement within the standard deviation about the
mean, they are indicative of a possible trend similar to what was
found radially. We see some sharp variations in the NE (0$^\circ$ to
90$^\circ$), SE (90$^\circ$ to 180$^\circ$) and SW regions compared
with the NW (270$^\circ$ to 360$^\circ$) region where the variation is
smooth. The plot is indicative of an azimuthal MG, with more
variations in the East compared with the West. This is possibly
related to the effect of an non-axisymmetric bar in the
LMC. \cite{Grady+2021ApJ} determined the metallicity profile along
directions aligned with the projected major and minor axes of the bar
in their fig.~11. That figure shows a flat gradient in the centre and
a negative MG with increasing distance from the central regions, with
strong asymmetries. Thus, our derived asymmetry in the radial MG of
the LMC is consistent with \cite{Grady+2021ApJ}.


\subsubsection{Interpretation of metallicity gradient}

\cite{Mazzi+2021} studied the SFH of the LMC using VMC data over
  an area of 96 deg$^2$, twice larger than
  \cite{Harris&Zaritsky2009AJ} who used the MCPS survey data. The SFH
derived by the former authors is similar to that of
\cite{Harris&Zaritsky2009AJ}, but presents a lower star-formation rate
(SFR) at young ages. Their maps show that the SFR at ages younger than
63~Myr is patchy and centralised. At older ages ($<$ 1.6~Gyr), the SFR
extends to larger radii. The SFR is concentrated in the bar region and
in three well-defined spiral arms. As we move to ages older than
1.6~Gyr, the bar and spiral features become less prominent and
eventually the SFH appears more like a circular distribution. The
period of enhanced star formation in the LMC occurs approximately
between 500 Myr and 4~Gyr, with two peaks of enhanced SFR: in the age
intervals 630~Myr to 1~Gyr and 1.6--4~Gyr (where the former is
associated with regions in the LMC bar). The RGB and RC star
contributions to the SFH are for ages older than 1.6~Gyr. Since we are
using the RGBs as our tracer, our radial MG trend is possibly
representative of a population with ages $\geq$ 1--1.6~Gyr. \cite{Mazzi+2021} report no significant MG in the LMC and
retrieve only a scattered distribution of metallicity that deviates
from the initially assumed age--metallicity relation. The authors
associate this with the relatively lower sensitivity of NIR bands to
variations in metallicity compared with optical passbands. According
to \cite{Mazzi+2021} the dependence on metallicity is related to
subtle changes in the mean RGB slope, in the position and shape of the
RC, and in the mean colour (compared with the RGB) and slope of the
main sequence. The negligible radial MG estimated in this study
using the RGB slope possibly reflects the interpretation of
\cite{Mazzi+2021}.

The metal-enhanced bar highlighted by this study and by previous
authors \citep[C16,][]{Grady+2021ApJ} is consistent with the findings
of \cite{Fraser-McKelvie+2019MNRAS}. These authors studied a sample of
128 barred galaxies and found that both the age and metallicity
gradients are flatter in the bar as opposed to the discs of their
galaxies. The presence of a bar appears to flatten or even erase
abundance gradients \citep{Alloin+1981A&Athemild}, probably by
inducing non-circular motions in the gas of the disc. Our results could
indicate that bars are efficient in radially mixing their stellar
populations even in a dwarf, one-armed spiral galaxy like the LMC,
just like it does in early- and late-type barred spiral galaxies
\citep{Seidel+2016MNRAS}. \cite{Bekki&Chiba2005} investigated the
  evolution of the LMC's MG using chemo-dynamical simulations. Their
  fig.~10 shows a steep gradient out to 3~kpc and a flattening at
  larger radii. Since a steep initial gradient is assumed in their
  simulation, the tidal interaction with the SMC and the MW, and the
  dynamical action of the LMC bar flattened the original MG in the
  model. Their steeper gradient in the central region is more
  consistent with the MCPS results of C16, whereas the flattened MG in
  the outer parts in more consistent with the present study. Our
  result implies that dynamical mixing of stellar populations with
  different metallicities by the LMC bar is perhaps more efficient
  than predicted by the simulations of \cite{Bekki&Chiba2005}.

\cite{Magrini+2016} found that the radial migration of stars can
contribute to a flattening of the MG in the outer parts of spiral
galaxies (e.g., M31, M33). The effect of stellar migration on the
metallicity distribution is also observed in the MW
\citep{Loebman+2016ApJ}. The flattening of the MG noticed in the LMC's
outer regions is possibly suggestive of a similar effect, whereby the
metal-rich stellar populations from the centre migrate outwards. There
could be many reason behind radial migration, e.g., transient spirals,
mergers and interaction with satellite galaxies, or interactions
between bars and spiral arms. The actual reason behind this phenomenon
in the LMC needs further verification, which is beyond the scope of
this work. The radially asymmetric MG in the LMC suggests a
  relatively flat trend on the western side, in the direction of the
  Bridge and the SMC, compared with the eastern side. The SW portion
  of the LMC's disc is known to be warped (by $\sim$4~kpc;
  \citealt{Choi+2018ApJ}) in the direction of the SMC because of their
  interaction history. The relatively metal-poor metallicity trend in
  the SW LMC hints at population mixing in this region, similar to the
  eastern SMC (C20), stemming from the LMC and SMC's tidal
  interaction.

\section{Summary}
We have successfully extended our technique of combining large-scale
photometric and spectroscopic data to estimate NIR metallicity maps of
the LMC. The results can be summarised as follows:
\begin{enumerate}
\item Our NIR metallicity maps exceed the previously obtained
  metallicity maps of C16 in terms of area coverage (three times
  larger), revealing trends across 105~deg$^2$ of the LMC. The
    spatial resolution of the NIR maps ranges between 50.37$\times$69.68~pc$^2$ and 201.79$\times$278.73~pc$^2$ for the
    innermost and outermost subregions, respectively.
\item We estimated RGB slopes in the $Y$ versus $Y-K_{\rm s}$ CMD of
  $\sim$4663 subregions within a radius of 6~kpc and converted the
  slopes to metallicity values using spectroscopic data of field RGs.
\item The mean metallicity of the LMC based on VMC data is
  $-0.42\pm0.04$~dex out to a radius of 6~kpc. This agrees well with
  the mean metallicities estimated in previous photometric and
  spectroscopic studies of RGs.
\item The RGB population drawn from a spatially homogeneous large-area
  photometric data set shows the existence of a negligible MG ($-0.008
  \pm 0.001$ dex kpc$^{-1}$) within 6~kpc. The estimated gradient is
  shallow compared with previous large-area photometric MG estimates
  \citep[C16,][]{Cioni2009A&Athemetallicity, Grady+2021ApJ}. Our NIR
  metallicity maps are possibly tracing a stellar population that is
  older than 1--1.6 Gyr.
\item Our results indicate that the LMC bar has played a role in
  turning the central region metal-rich and flattening the gradient
  within the central 3~kpc. The trend of a flattened gradient in the outer LMC is possibly caused by mixing induced by the LMC bar or by radial migration of the metal-rich population from the centre outwards.
\item The LMC has an asymmetric radial MG, possibly caused by the
  non-axisymmetric bar. The eastern disc has a steeper metallicity
  trend compared with the western disc. The metal-poor and flattened
  trend in the SW, in the direction of the Bridge and the SMC, is
  possibly analogous to a similar trend noted in the eastern SMC by
  C20. We suspect that this could be a result of the tidal interaction
  in the LMC--SMC system.
\item Our study supports further spectroscopic investigations of the
  LMC using large-scale multi-fibre spectrographs like the 4m
  Multi-Object Spectroscopic Telescope
  \citep[4MOST;][]{deJong+2014SPIE, deJong2019Msngr}. Detailed studies
  of individual chemical abundances and chemo-kinematics for regions
  like the SW quadrant of the LMC and the eastern part of the SMC may
  yield important clues as to the LMC--SMC interaction history.

\end{enumerate}

\section*{Acknowledgements}
We thank the Cambridge Astronomy Survey Unit (CASU) and the Wide Field
Astronomy Unit (WFAU) in Edinburgh for providing the necessary data
products under the support of the Science and Technology Facilities
Council (STFC) in the UK. This study was based on observations made
with VISTA at the La Silla Paranal Observatory under programme ID
179.B-2003. This project has received funding from the European
Research Council (ERC) under the European Union's Horizon 2020
research and innovation programme (grant agreement 682115). We
  acknowledge Stefano Rubele for generating the PSF photometry
  catalogue used in this work. We are grateful for the use of the
  computer cluster at the University of Hertfordshire, to Simone
  Zaggia for making available local computers at the University of
  Padova, and acknowledge the ERC Consolidator Grant funding scheme
  (project STARKEY, grant agreement 615604) for supporting S.
  Rubele's work. We thank Andrew A. Cole for providing CaT
  spectroscopic data of field RGs in the LMC which helped in
  estimating the slope--metallicity calibration relation. We also
  thank Martin A. T. Groenewegen for his comments which helped improve
  the manuscript. We thank the anonymous referee for their suggestions which helped improve the clarity of the manuscript.

\section*{Data Availability}
The PSF catalogue of the LMC used in our work is proprietary to
  the VMC team, with a data release planned in mid 2022.

\bibliographystyle{mnras}
\bibliography{bibliography}








\bsp	
\label{lastpage}
\end{document}